\documentclass{article}
\usepackage[english]{babel}
\usepackage[utf8]{inputenc}
\usepackage{graphicx}
\usepackage{float}
\usepackage{geometry}
\usepackage{placeins}
\usepackage{braket}
\usepackage{amsmath}
\usepackage{amsfonts}
\usepackage{bm}
\usepackage{soul}
\usepackage{xcolor}
\usepackage{esvect}
\usepackage{tabularx}
\usepackage{multirow}
\usepackage{outlines}
\usepackage{subfig}
\usepackage[hidelinks]{hyperref}
\usepackage{authblk}
\usepackage[backend=bibtex,giveninits=true,style=nature]{biblatex}
\usepackage{numprint}
\usepackage[makeroom]{cancel}

\npthousandsep{,}

\DeclareMathOperator*{\argmin}{argmin}

\newcommand{\nel}{n_{\mathrm{el}}}
\newcommand{\nnuc}{n_{\mathrm{nuc}}}
\newcommand{\nsu}{n_{\uparrow}}
\newcommand{\nsd}{n_{\downarrow}}
\newcommand{\ndet}{n_{\mathrm{det}}}

\newcommand{\nrbf}{n_{\mathrm{rbf}}}
\newcommand{\nemb}{n_{\mathrm{emb}}}
\newcommand{\bmrprop}{\bm{r}^{\text{prop}}}
\newcommand{\rprop}{r^{\text{prop}}}

\newcommand{\thetash}{\theta^{\textrm{sh}}}

\title{Solving the electronic Schrödinger equation for multiple nuclear geometries with weight-sharing deep neural networks}
\author[$\dagger$,*]{Michael Scherbela}
\author[$\ddagger$,a,*]{Rafael Reisenhofer}
\author[$\dagger$,*]{Leon Gerard}
\author[$\dagger$,$\mathsection$]{Philipp Marquetand}
\author[$\dagger$,$\ddagger$,$\mathparagraph$]{Philipp Grohs}

\affil[$\dagger$]{Research Network Data Science @ Uni Vienna, Kolingasse 14-16, A-1090 Vienna, Austria}
\affil[$\ddagger$]{Faculty of Mathematics, University of Vienna, Oskar-Morgenstern-Platz 1, A-1090 Vienna, Austria}
\affil[$\mathsection$]{Faculty of Chemistry, University of Vienna, Währinger Straße 17, 1090 Vienna, Austria}
\affil[$\mathparagraph$]{Johann Radon Institute for Computational and Applied Mathematics,  Austrian Academy of Sciences, Altenbergerstrasse 69, 4040 Linz, Austria}
\affil[a]{\href{mailto:rafael.reisenhofer@univie.ac.at}{reisenhofer@uni-bremen.de}}
\affil[*]{These authors contributed equally}

\date{}   
\begin{document}

\maketitle

\begin{abstract}
Accurate numerical solutions for the Schrödinger equation are of utmost importance in quantum chemistry. However, the computational cost of current high-accuracy methods scales poorly with the number of interacting particles. Combining Monte Carlo methods with unsupervised training of neural networks has recently been proposed as a promising approach to overcome the curse of dimensionality in this setting and to obtain accurate wavefunctions for individual molecules at a moderately scaling computational cost. These methods currently do not exploit the regularity exhibited by wavefunctions with respect to their molecular geometries.
Inspired by recent successful applications of deep transfer learning in machine translation and computer vision tasks, we attempt to leverage this regularity by introducing a weight-sharing constraint when optimizing neural network-based models for different molecular geometries. That is, we restrict the optimization process such that up to 95 percent of weights in a neural network model are in fact equal across varying molecular geometries. We find that this technique can accelerate optimization when considering sets of nuclear geometries of the same molecule by an order of magnitude and that it opens a promising route towards pre-trained neural network wavefunctions that yield high accuracy even across different molecules.
\end{abstract}
\section{Introduction}
%
Using a deep neural network-based ansatz for variational Monte Carlo (VMC) has recently emerged as a novel approach for highly accurate ab-initio solutions to the multi-electron Schrödinger equation \cite{Han_2019, Hermann2020, Manzhos_2020, FermiNet, FermiNet_DMC}. It has been observed that such methods can exceed gold-standard quantum-chemistry methods like CCSD(T) \cite{Bartlett_CoupledCluster} in accuracy, with a computational cost per step scaling only with $\mathcal{O}(N^4)$ in the number of electrons \cite{FermiNet}. This suggests a drastic improvement from classical quantum-chemistry methods such as CCSD(T), or CISDTQ\footnote{Configuration interaction singles, doubles, triples, quadruples}, which scale with $\mathcal{O}(N^7)$ and $\mathcal{O}(N^{10})$, respectively. However, due to the large number of free parameters and the need for Monte Carlo integration, the constant prefactor for neural network-based methods is typically much larger than for classical approaches such that even systems of modest size still require days or weeks of computation when using highly optimized implementations on state-of-the-art hardware \cite{FermiNet2}. This often renders deep neural network (DNN)-based ansatz methods unfeasible in practice, in particular when highly accurate results for a large number of molecular geometries are required.

%
Among such tasks are computational structure search, determination of chemical transition states, and the generation of training datasets for supervised machine learning algorithms in quantum chemistry. Latter methods are applied with great success to interpolate results of established quantum chemistry methods such as energies and forces \cite{Unke2021_MLForceFields,Behler2021CR, DeepMind21}, properties of excited states \cite{Westermayr2020_MLExcitedStates}, underlying objects such as orbital energies \cite{Schuett2019}, or the exchange energy \cite{Burke2020_ML_DFT_functional}. Given sufficient training data, these interpolations already achieve chemical accuracy relative to the training method (e.g. Density Functional Theory) \cite{Faber2017_MLLowPredictionError}, highlighting the need for increasingly accurate ab-initio methods which can be used to generate reference training data.

%
The goal of making DNN-based VMC applicable for the generation of such high-quality datasets for previously untractable molecules is a key motivation for this work. The apparent success of supervised learning in quantum chemistry suggests a high degree of regularity of the aforementioned properties and the wavefunction itself within the space of molecular geometries.

%
Here, we aim to exploit potential regularities of the wavefunction within the space of molecular geometries already during VMC optimization by applying a simple technique called weight-sharing. Throughout optimizing instances of the same neural network-based wavefunction model for different molecular geometries, we enforce that for large parts of the model, each instance has the exact same neural network weights. In particular, this means that on the parts of the model where weight-sharing is applied, each instance computes precisely the same function. We note that this idea is reminiscent of (and inspired by) the ML technique \emph{deep transfer learning} where parts of a pre-trained model are reused for different similar tasks and which has led to breakthrough results, for example in natural language processing \cite{transfer2} or computer vision \cite{transfer1}.

%
Weight-sharing can be viewed as a regularization technique which requires large parts of the optimized model to work equally well for a potentially wide variety of different nuclear, or even molecular, geometries. Under the assumption that the wavefunctions are sufficiently regular across geometries, it should therefore have a stabilizing effect on the optimization process and yield wavefunctions that also generalize well to new molecular geometries when used as an initial guess before optimization. For a shared weight, each gradient descent update during optimization for a specific geometry is applied to the complete set of considered geometries. Weight-sharing therefore has the potential of significantly accelerating the optimization process.

%
Our main numerical results highlight the benefits of weight-sharing as compared to independent optimization (Sec.~\ref{subsec:sharedopt}) and the applicability of pre-trained shared weights for new calculations (Sec.~\ref{subsec:reuse}). In particular, we show that by applying these techniques in combination with second-order optimization, it is possible to consistently reach the energies of MRCI-F12 reference calculations -- up to chemical accuracy -- for molecules up to the size of ethene after only $\mathcal{O}(10^2)$ optimization epochs per geometry. Note that for most recently proposed DNN-based VMC methods, wavefunctions are typically being optimized for $\mathcal{O}(10^4)$ - $\mathcal{O}(10^5)$ epochs. To further demonstrate the applicability of the proposed framework in practice, we calculate the transition path for H$_4^+$ between two symmetry equivalent minima, wavefunctions for a set of differently twisted and stretched ethene configurations, as well as the potential energy surface (PES) -- including forces -- of a H$_{10}$ chain on a 2D grid of nuclear coordinates (Sec.~\ref{subsec:applications}). While nuclear forces are also available for other methods, such as domain-based local pair natural orbital (DLPNO)-Coupled-Cluster \cite{Matthews2020_Forces_CoupledCluster}, they are not available for all types of systems \cite{Liakos2020_CC_Benchmark}, or come at a significant additional computational cost, while our approach yields forces at a low incremental cost.
\section{Results}
%
%
To investigate weight-sharing for neural network-based models in VMC, we consider a framework, dubbed DeepErwin, where the trial wavefunction is modeled similar to the recently proposed PauliNet \cite{Hermann2020}, with modifications leading to an overall smaller network that yields higher accuracies. The basic idea behind this model is to enhance a Slater determinant ansatz with deep neural networks, where initial orbitals are obtained from a CASSCF (complete active space self-consistent field) calculation with a small basis set and a small active space. The resulting wavefunction is then modified by applying a backflow transformation to the orbitals as well as to the electron coordinates, and via an additional Jastrow factor. All these enhancements depend on an embedding of the electron coordinates into a high-dimensional feature space which takes into account interactions with all other particles. This embedding is based on Electronic SchNet \cite{NIPS2017_303ed4c6} and PauliNet. Cusp correction is performed explicitly \cite{Ma2005_CuspScheme}. In contrast to PauliNet, we use an additional equivariant backflow shift for the electron coordinates, smaller embedding networks, and no residual in the embedding layers.
For a realization of the wavefunction model implemented in the DeepErwin framework, the energy can be approximated through Monte Carlo integration. To eventually obtain the wavefunction of the ground state for a specific molecule, this energy is minimized by applying gradient descent steps to the free parameters of the wavefunction model. A detailed description of our architecture and the optimization procedure can be found in the methods section.

\begin{figure}[!tbp]
\centering
\subfloat[]{
    \label{fig:overview_shared_parts_nn}
    \includegraphics[height=4.8cm]{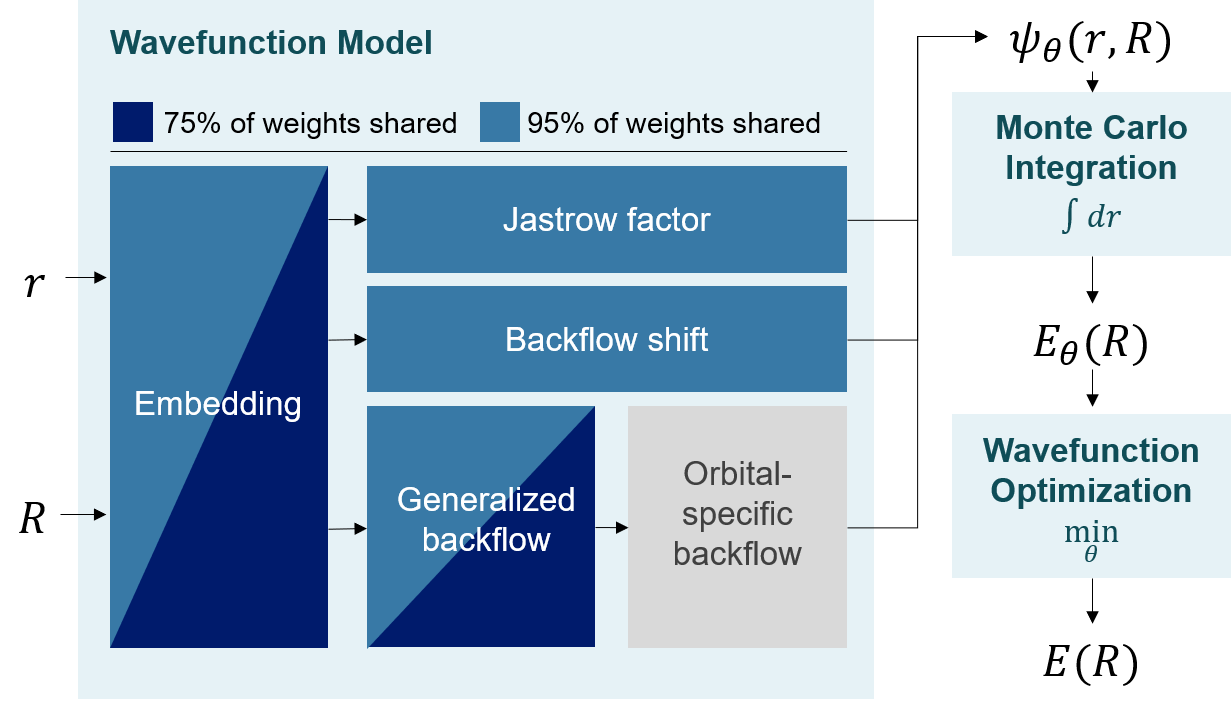}}\hspace{.2cm}
    \subfloat[]{\label{fig:accuracy_comparison}
     \includegraphics[height=6cm]{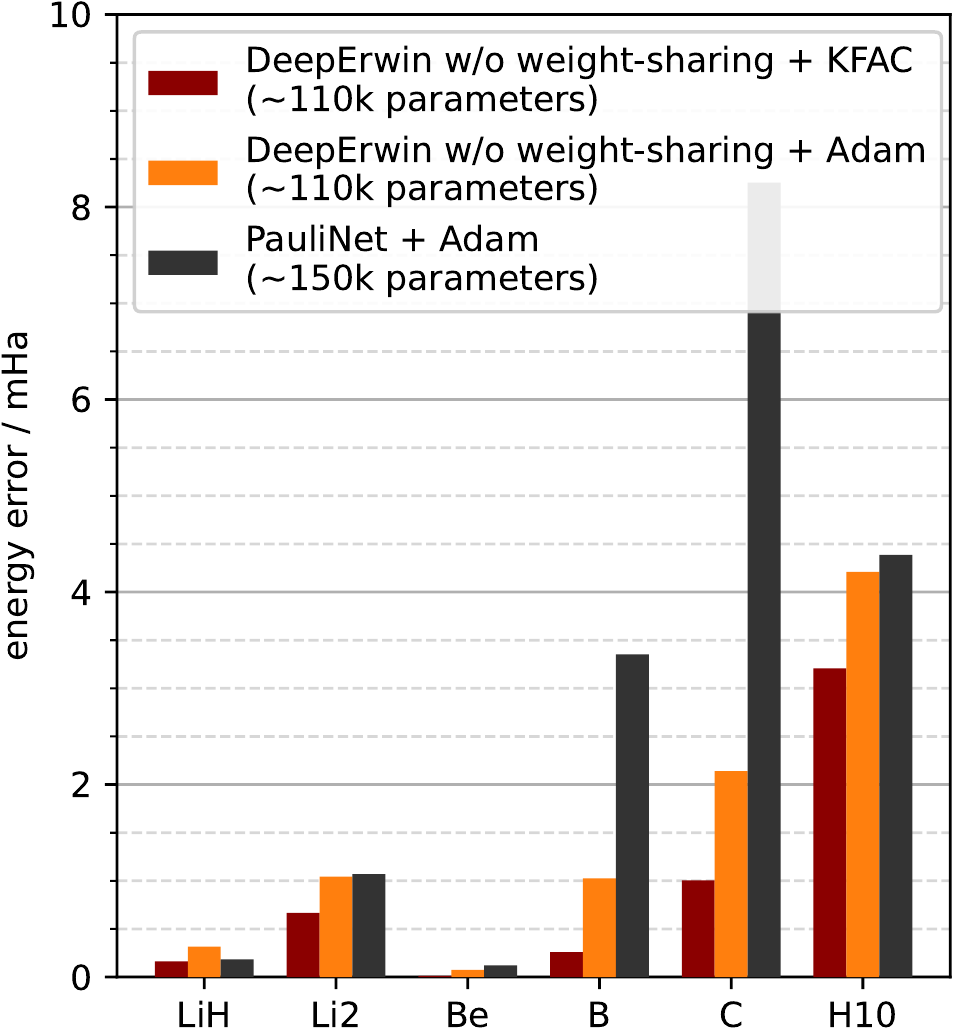}}
    \caption{(a) Overview of the neural network parts of the wavefunction model implemented in the DeepErwin framework, including a top-level visualization of the two weight-sharing setups considered in our experiments. (b) Energy of optimized wavefunctions relative to reference calculations \cite{FermiNet} for DeepErwin with first-order optimization (ADAM), second-order optimization (K-FAC), and PauliNet \cite{Hermann2020}. DeepErwin baseline solutions without weight-sharing significantly outperform PauliNet across the tested systems despite a smaller number of parameters.}
\end{figure}

%
A top-level overview of the neural network parts of DeepErwin is shown in Figure~\ref{fig:overview_shared_parts_nn}. In comparison with PauliNet, we find that, without using weight-sharing constraints, we achieve superior results with significantly fewer trainable weights for the small systems tested in \cite{Hermann2020} when training for the same number of epochs (see Fig.~\ref{fig:accuracy_comparison}). This indicates that the architecture implemented in DeepErwin provides a meaningful baseline to investigate the effects of weight-sharing on the optimization process. The application of weight-sharing is of course not limited to this specific model but could equally well be adapted for any neural network-based wavefunction model.

All subsequently reported results were obtained via second-order optimization using the Kronecker-factored approximate curvature (K-FAC) \cite{martens2015optimizing} which was already implemented for FermiNet \cite{FermiNet}. To show that our findings are consistent across different types of optimization, we also report results for the experiments from Sec.~\ref{subsec:sharedopt} and Sec.~\ref{subsec:reuse} when using the well-known ADAM algorithm \cite{Adam} in the supplementary materials.
\subsection{Accelerated optimization through weight-sharing}
\label{subsec:sharedopt}
\begin{figure}[!tbp]
    \centering
    \includegraphics[width=\columnwidth]{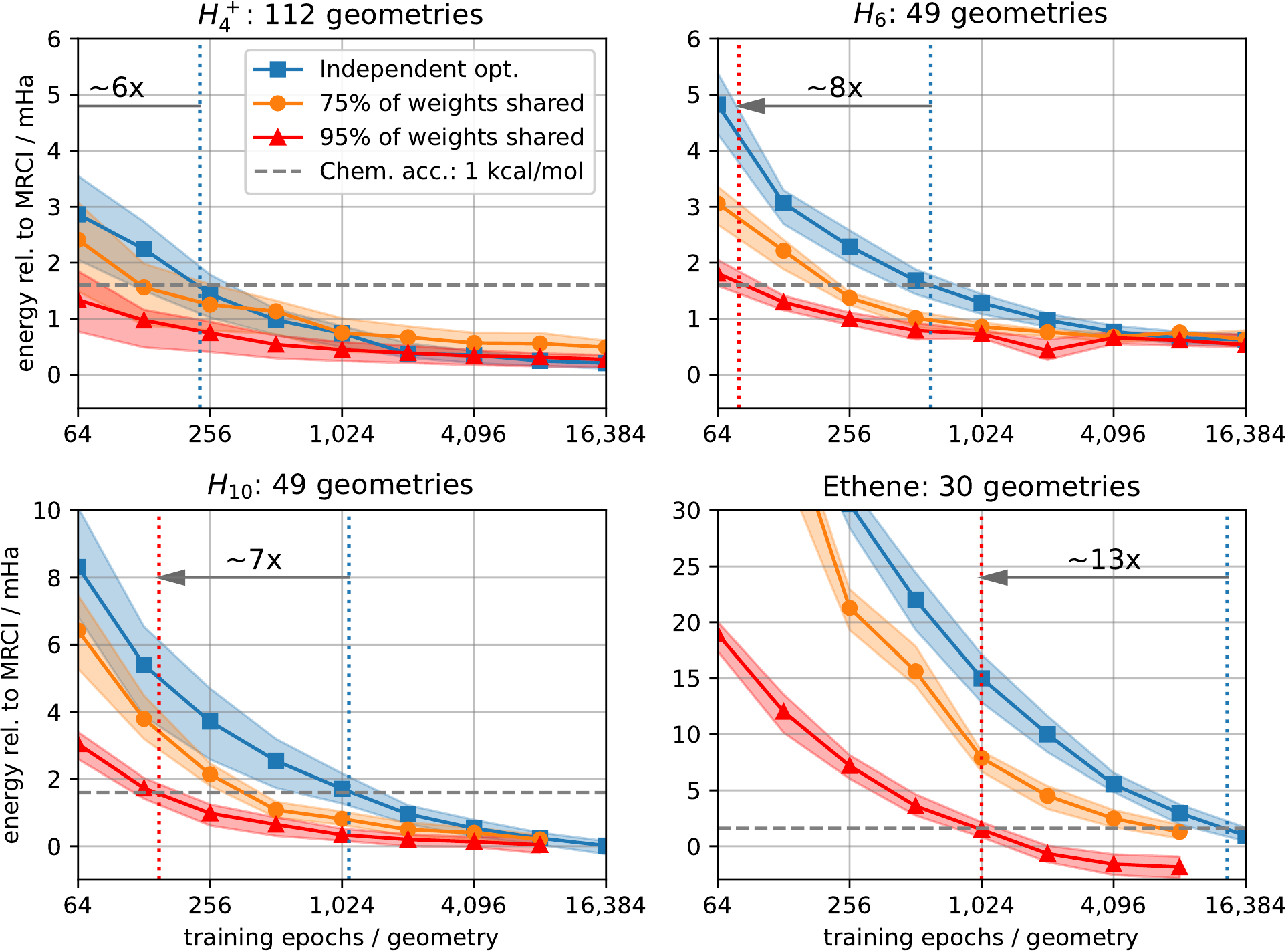}
    \caption{Mean evaluation error relative to the reference calculation (MRCI-F12(Q)/cc-pVQZ-F12) as a function of training epochs per geometry for four different sets of nuclear geometries. Shadings indicate the area between the 25\,\% and the 75\,\% percentile of considered geometries. For each method, we plot intermediary and final results for optimizations that ran for a total number of \numprint{16384} epochs per geometry, respectively \numprint{8192} epochs per geometry for shared optimization of H$_{10}$ and ethene. Note that for H$_{10}$, the reference energies here differ slightly from the reference energy in Figure~\ref{fig:accuracy_comparison}, which was not available for the complete set of 49 geometries.} 
    \label{fig:parallel_speedup}
\end{figure}
%
%
The implemented architecture as well as the hyperparameters used in our experiments are designed to allow for a maximum number of free parameters to be potentially shared across geometries. Two parts of the model that should be particularly well-suited for weight-sharing are the electron coordinate embedding and the generalized part of the backflow factor. Both basically serve as feature extractors that compute high-dimensional embeddings of electron coordinates and are therefore not necessarily required to perform geometry-specific computations. The orbital-specific part of the backflow factor, on the other hand, has a one-to-one correspondence to the orbitals yielded by CASSCF for a given geometry, suggesting that the neural network weights defining it cannot be shared across geometries in a meaningful way.

%
When weight-sharing is restricted to the embedding and the generalized part of the backflow factor, usually about 75\,\% of the weights in the model are covered by weight-sharing constraints. In the most extreme case, when all weights are being shared -- except the ones defining the orbital-specific backflow -- this number grows to roughly 95\,\% (cf. Fig.~\ref{fig:overview_shared_parts_nn}). Note that precise counts for the numbers of total and shared model parameters used for the experiments in this section slightly differ between different molecules (see Table~\ref{table:parameternumbers}).

We evaluate both these setups in four cases by computing the PES of four different molecules, namely H$_4^+$, the linear hydrogen chains H$_6$ and H$_{10}$, and ethene. For H$_4^+$, we consider a diverse set of 112 different configurations, covering both low-energy relaxed geometries, as well as strongly distorted configurations. For both hydrogen chains, we calculate the wavefunction of the ground state for 49 different nuclear geometries that lie on a regular grid with respect to a parametrization based on the distance $a$ of two adjacent H-atoms and the distance $x$ between these H$_2$ pairs. In the case of twisted ethene, the set of 30 geometries iterates over ten different twist angles and three different bond lengths for the carbon-carbon double bond. Figure~\ref{fig:H10_PES_0} depicts the resulting PES for H$_{10}$. Figure~\ref{fig:Barrier_kfac_1} plots the obtained energies for the minimum-energy-path from the non-twisted equilibrium geometry to the 90$^\circ$ rotated molecule, considering for each twist angle the CC bond-length with the lowest energy.

For all four molecules, we compare the optimization of the wavefunction models when applying weight-sharing with the respective independent optimizations. The results of these experiments are compiled in Figure~\ref{fig:parallel_speedup}. Across all physical systems, the optimization converges fastest when 95\,\% of the weights are being shared. In particular, we find that in this case, the reference energy (MRCI-F12(Q)/cc-pVQZ-F12; see Methods section for computational details) can be reached, up to chemical accuracy, between 6 and 13 times faster than when optimizing the respective geometries independently of each other.

To test our findings in this section against a more difficult benchmark, we also performed an additional experiment where independent optimizations for different ethene configurations were fully initialized with weights from a wavefunction that had already been optimized for a similar but different molecular configuration using an independent optimization scheme. While this approach seems to be advantageous during early optimization as compared to a scheme that applies a weight-sharing constraint for 95\,\% of the weights in the model, at the time the wavefunctions reach chemical accuracy, shared optimization without pre-training outperforms this new baseline to a degree comparable with the results shown in Figure~\ref{fig:parallel_speedup}. Detailed results for this experiment can be found in Figure~S3 in the supplementary materials.
\subsection{Shared optimization as pre-training}
\label{subsec:reuse}
\begin{figure}[!tbp]
    \centering
    \includegraphics[width=\columnwidth]{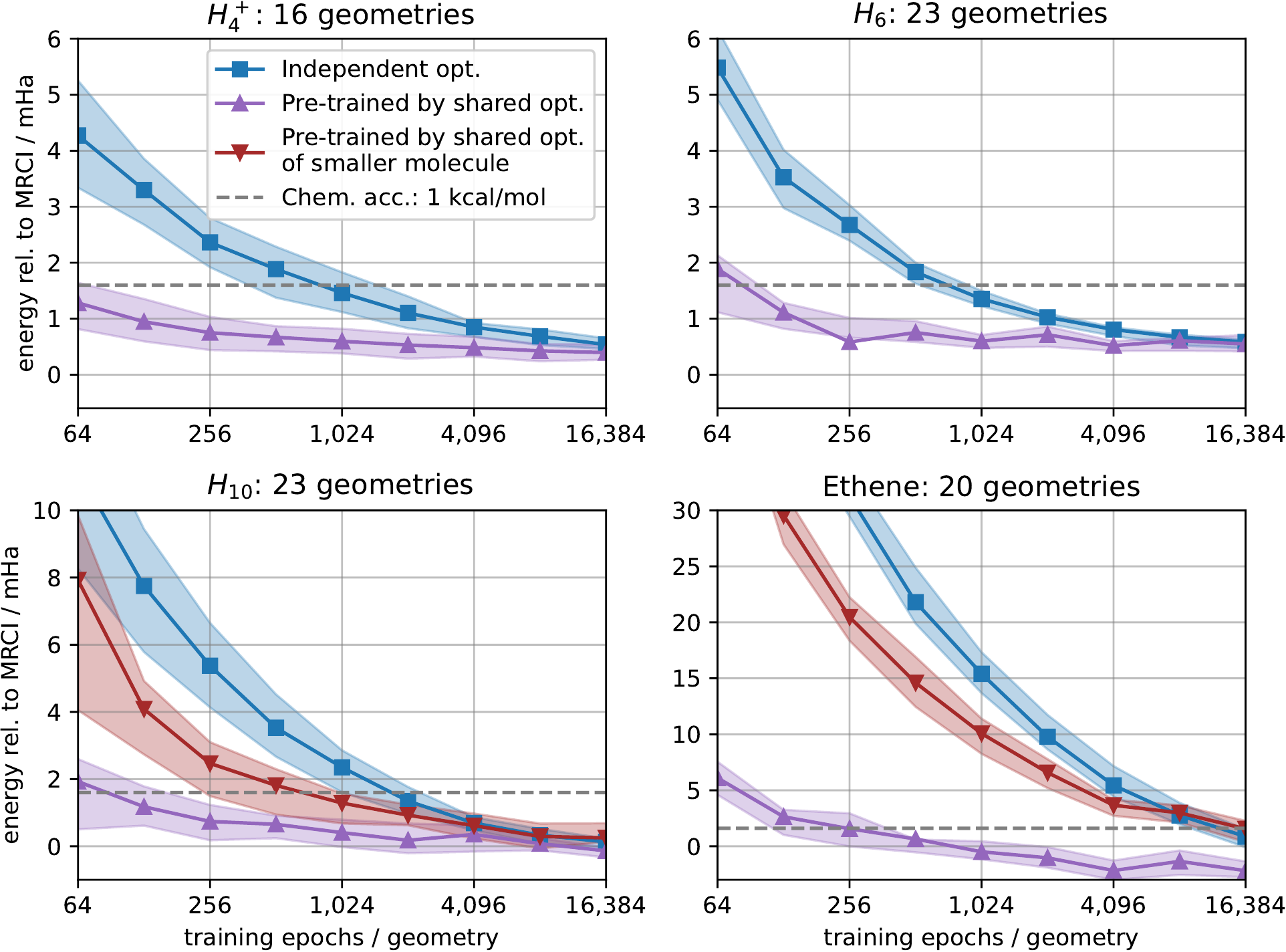}
    \caption{Mean evaluation error relative to a reference calculation (MRCI-F12(Q)/cc-pVQZ-F12 (Molpro)) as a function of training epochs per geometry for four different sets of nuclear geometries. Plots show energy errors for two different pre-training schemes, compared to a standard non-pretrained optimization: Pre-training on different geometries of the same molecules, and pre-training on geometries of a smaller molecule (methane for ethene and H$_6$ for H$_{10}$). Shadings indicate the area between the 25\,\% and the 75\,\% percentile of considered geometries. For each method, we plot intermediary and final results for independent optimizations that ran for a total number of \numprint{16384} epochs.}
    \label{fig:reuse_speedup}
\end{figure}
Results from the previous section show that even in a setting where ground state wavefunctions are being closely approximated for a wide range of nuclear geometries by different instances of our model, the overwhelming number of free parameters can in fact be identical across those instances. This suggests that parts of our model that were successfully optimized using a weight-sharing constraint encode general computational building blocks that are highly suitable for the approximation of different wavefunctions. In particular, one could hope that those building blocks also generalize well to molecular geometries for which they were not previously optimized.

To test this hypothesis, we consider for each of the four molecules from the previous experiment a small new set of nuclear geometries that were not part of the original shared optimization. For these sets, we compare two types of independent optimization. In one case, a default random method is used to initialize the weights of the neural networks before optimization. In the second case, we reuse results from the previous optimization in the sense that all weights of the model that were shared in the first experiment are now initialized from the result of this optimization. For the remaining 5\,\% of weights, default random initialization is applied.

Pushing this approach even further, we also use previously optimized shared weights to initialize wavefunction models for an entirely different molecule. In particular, we use shared weights that were optimized for the hydrogen chain H$_6$ to initialize models for H$_{10}$, and weights that were optmized for methane to initialize models for ethene. This is possible because the embedding network architecture is independent of the number of particles in the respective molecule. If successful, this method can be used to pre-train weights for large and expensive molecules by solving the Schrödinger equation for smaller, computationally cheaper systems.

The results of these experiments are shown in Figure~\ref{fig:reuse_speedup}. For all four considered molecules, we used weights that were optimized with a shared-weight constraint on a set of different geometries (cf. Fig.~\ref{fig:parallel_speedup}) for \numprint{8192} optimization epochs per geometry. Across all systems, pre-training via shared optimization with different geometries of the same molecules dramatically accelerates the subsequent optimization such that the reference energy can be consistently reached up to chemical accuracy after little more than a hundred optimization epochs. In the case of H$_{10}$, the usage of weights that were pre-trained on different configurations of a smaller molecule also yields significant, albeit much smaller improvements. When using methane configurations to pre-train a wavefunction model for ethene, however, we could only find slight improvements during early optimization.
\subsection{Calculating transition paths and forces}
\label{subsec:applications}
The significant speed-ups obtained through weight-sharing enable efficient computational studies for systems that consist of many different geometries of the same molecule. We demonstrate the capabilities of our approach on two exemplary tasks: Finding transition paths and calculating potential energy surfaces.
As a first example, we calculate the transition path for H$_4^+$ between two symmetry equivalent minima via a specific transition state previously proposed in the literature \cite{Alijah2008_H4plus}. For all 19 points along the transition path, wavefunctions are optimized simultaneously for \numprint{7000} epochs per geometry using a weight-sharing constraint that covers about 95\,\% of total weights in the model. We furthermore compute the energies along a reaction path for twisted ethene which describes a rotation of the twist by 90$^{\circ}$. The ten geometries considered in this task are a subsample of the 30 geometries previously used in the computations shown in Fig.~\ref{fig:parallel_speedup}. As a baseline, we consider independent optimization without a weight-sharing constraint as well as classical methods from computational chemistry.

\begin{figure}[!tbp]
    \centering
    \subfloat[]{\label{fig:Barrier_kfac_0}\centering\includegraphics[height=7cm]{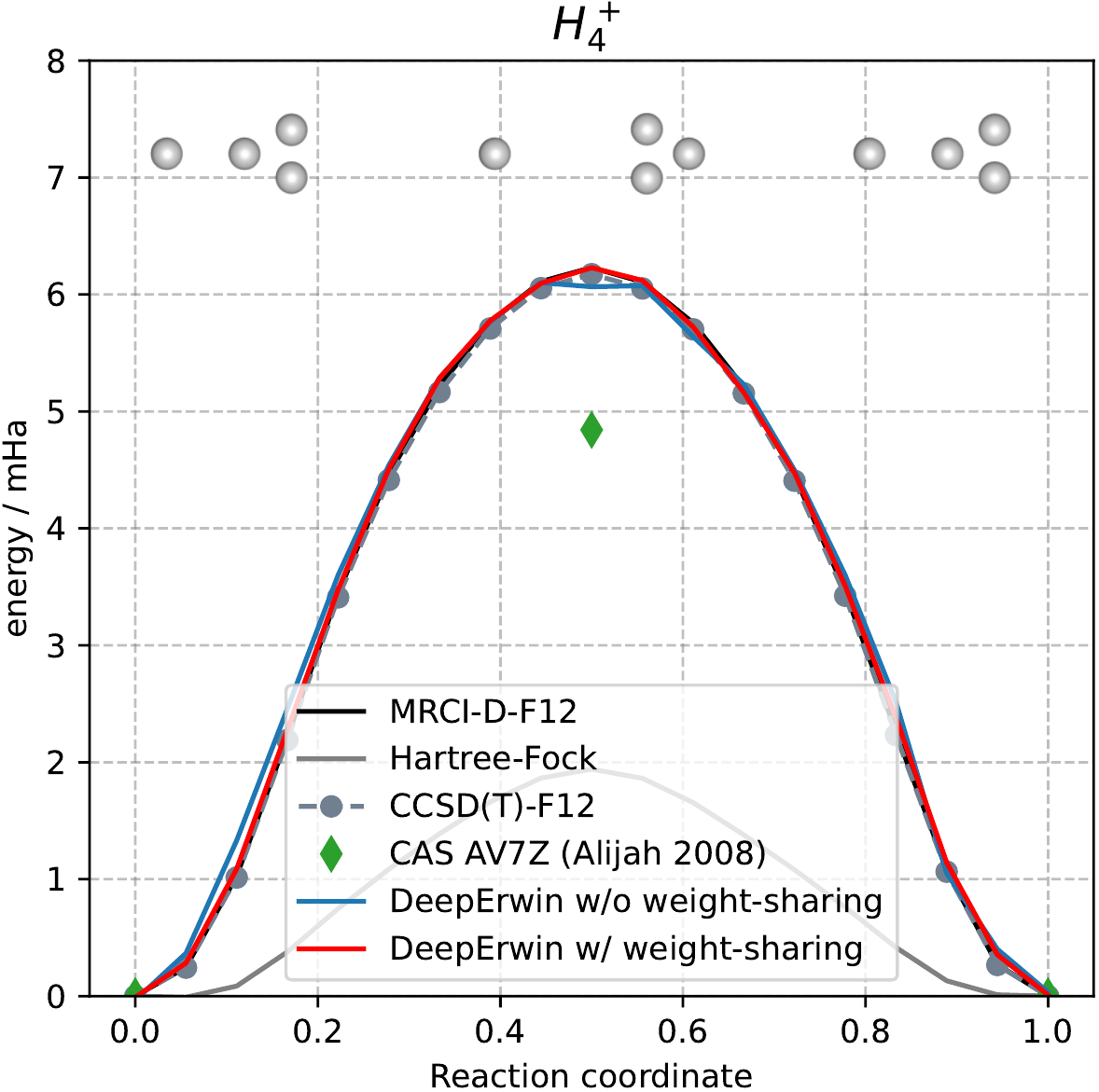}}\hspace{.2cm}
    \subfloat[]{\label{fig:Barrier_kfac_1}\centering\includegraphics[height=7cm]{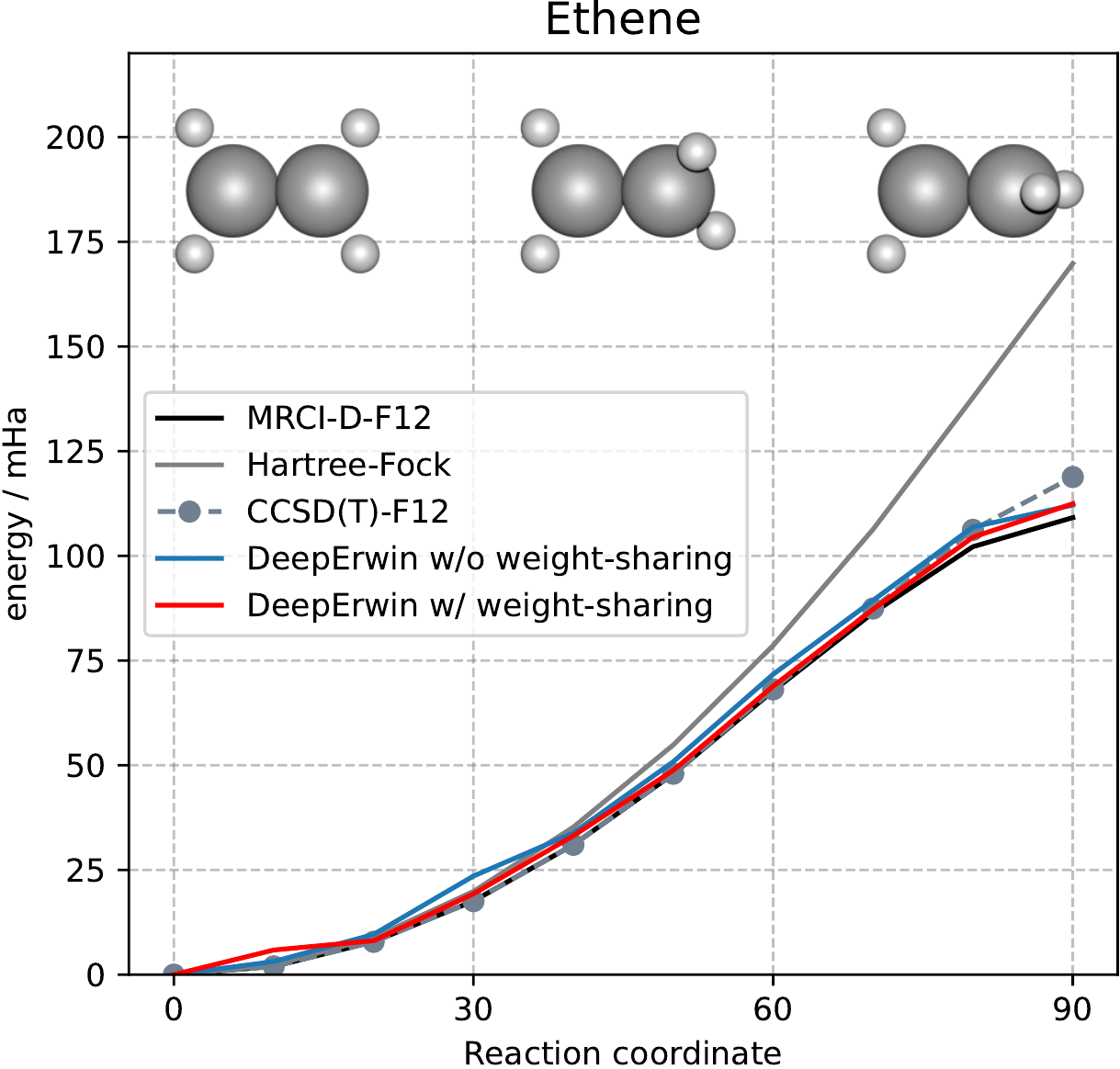}}
    \caption{(a) Energy of H$_4^+$ along reaction path between symmetry equivalent minima, via transition state 1 as defined in \cite{Alijah2008_H4plus}. DeepErwin with weight-sharing constraints is in perfect agreement with MRCI-F12(Q) and CCSD(T)-F12 after \numprint{7000} optimization epochs per geometry. (b) Energies for ten geometries that describe a rotation of the twist for twisted ethene by 90$^{\circ}$. Results for DeepErwin with and without weight-sharing are plotted after \numprint{8192} optimization epochs per geometry.}
    \label{fig:Barrier_kfac}
\end{figure}

The results of these calcualations are shown in Figure~\ref{fig:Barrier_kfac_0} and Figure~\ref{fig:Barrier_kfac_1}, respectively. We find our method to be in superior agreement with high-accuracy reference calculations: DeepErwin with weight-sharing predicts barrier heights that agree with MRCI within 1µHa (0.02\%) for H$_4^+$, and 3.3 mHa (3\%) for ethene. This leads us to believe that the barrier height for H$_4^+$ has been underestimated by approximately 1~mHa in previous high-accuracy calculations \cite{Alijah2008_H4plus}. For the electronically challenging case of twisted ethene, both Hartree-Fock as well as CCSD(T)-F12 overestimate the energy of the 90$^\circ$ twisted molecule. DeepErwin, however, yields barrier energies that are in much closer agreement with the MRCI-D-F12 calculations.

\begin{figure}[!tbp]
    \centering
    \subfloat[]{\label{fig:H10_PES_0}\centering
    \includegraphics[height=4.5cm]{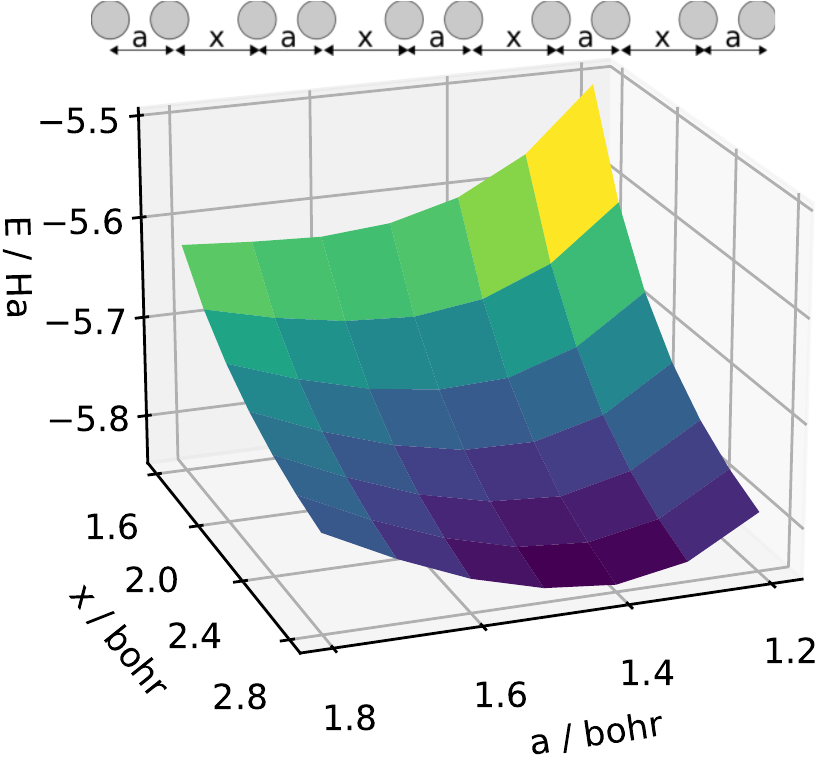}}\hspace{.2cm}
    \subfloat[]{\label{fig:H10_PES_1}\centering
    \includegraphics[height=5cm]{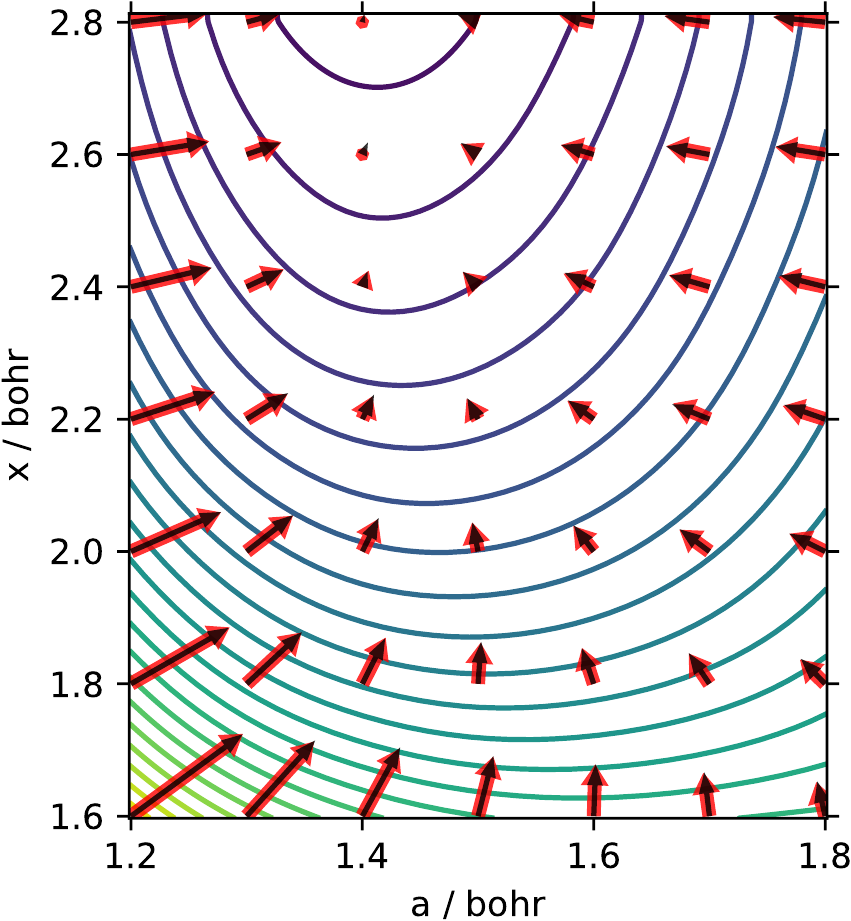}}\hspace{.2cm}
    \subfloat[]{\label{fig:H10_PES_2}\centering
    \includegraphics[height=5cm]{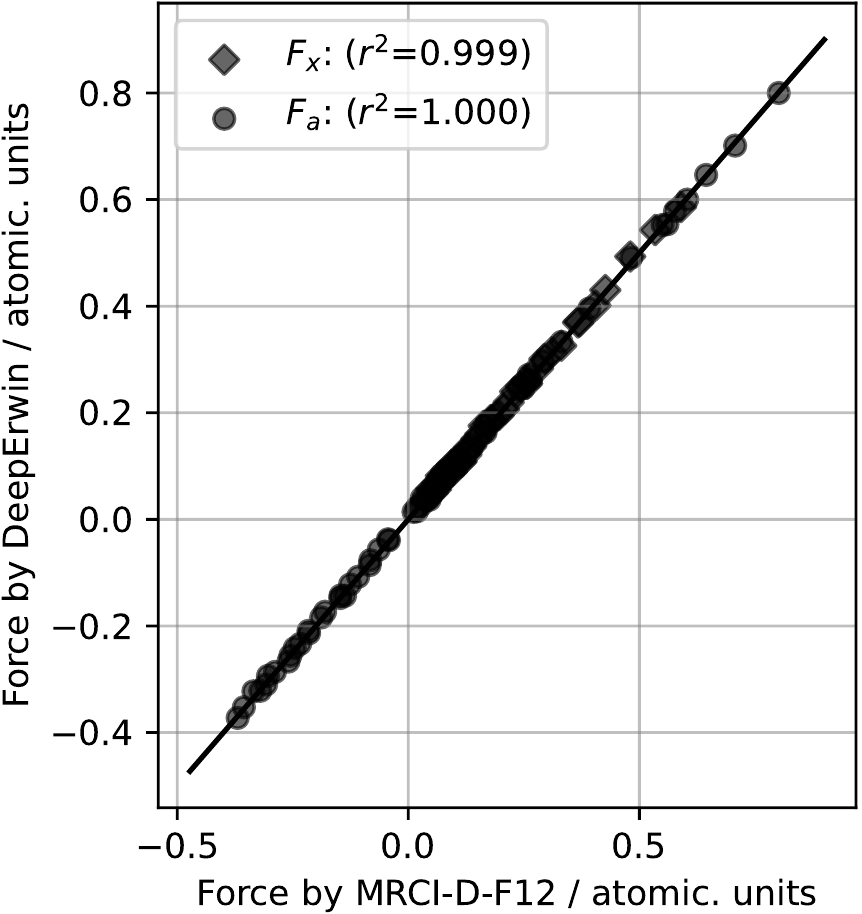}}
    \caption{(a) Potential energy surface (PES) for the H$_{10}$ chain. The variable $a$ is representing the distance between two H-atoms and $x$ is the distance between two H$_2$ molecules.  The lowest ground state of the PES describes the dimerization. (b) Cubic interpolation of the PES with force vectors for the H$_{10}$ chain. Red arrows depict force vectors computed by DeepErwin via the Hellmann-Feynman theorem, whereas black arrows represent numerical gradients that are based on finite differences of MRCI-F12 reference calculations. (c) Forces computed by DeepErwin plotted against the respective forces obtained from finite differences of the MRCI-F12 reference calculation.}
    \label{fig:H10_PES}
\end{figure}

For calculating the transition paths, we used predefined sets of geometries as a given input. In many cases, however, it is not a-priori clear which nuclear geometries are of interest in a given task, and a careful exploration of the respective PES is required. To do this efficiently, not only energies, but also forces on the nuclei, that is, gradients of the energy with respect to the nuclear coordinates, are required. For realizations of the DeepErwin wavefunction model, these forces can be calculated in a straightforward and computationally efficient fashion via the Hellman-Feynman theorem \cite{HellmanFeynman} and by applying established variance correction schemes to accelerate convergence of the Monte Carlo integration (see Methods section). The PES and the corresponding forces for the linear hydrogen chain H$_{10}$, evaluated on a regular grid of 49 geometries, are depicted in Figure~\ref{fig:H10_PES_0} and Figure~\ref{fig:H10_PES_1}, respectively. The corresponding wavefunctions were optimized for \numprint{8192} epochs per geometry using a weight-sharing constraint for approximately 95\,\% of the model weights.
We find that the energetic minimum is given by the dimerization into five H$_2$ molecules with a covalent bond of $a = 1.4$ Bohr each, rather than for an equally spaced arrangement of atoms. This is an instance of the well-known Peierls distortion \cite{peierls1955quantum}.

Figure~\ref{fig:H10_PES_2} shows that the force vectors obtained by DeepErwin via the Hellman-Feynman theorem are in perfect agreement with the forces computed from finite differences of MRCI-F12 reference calculations. Our computational experiments do not show any signs of spurious Pulay forces \cite{Pulay}, which occur when the approximated wavefunction is not an eigenfunction of the Hamiltonian. This suggests that DeepErwin yields not only highly accurate energies, but also highly accurate wavefunctions.
\section{Discussion}
%
Sharing the weights of realizations of a neural network-based wavefunction model across different nuclear geometries can significantly reduce the computational cost of VMC optimization. For several molecules, we found that weight-sharing can in fact decrease the number of required optimization epochs per geometry to reach a given reference energy up to chemical accuracy by a factor between 6 and 13. Our approach yields the best results when an overwhelming majority of network weights is being shared, suggesting a strong regularity of wavefunctions of the considered molecules within the space of nuclear geometries. Further evidence for this regularity is also provided by a concurrent method named PESNet, which was released shortly after DeepErwin was first available as a preprint \cite{gao2021ab}. PESNet is based on the FermiNet model and employs a meta graph neural network (GNN) to simultaneously learn wavefunctions for a complete potential energy surface (PES) such that after optimization, the model parameters for a new geometry can be predicted via a simple forward pass through the meta GNN. The fact that this approach reliably yields high-accuracy results for different configurations that were sampled from a PES further underlines our observation that the regularity of wavefunctions within the space of geometries can be heavily exploited for DNN-based QMC methods by also enforcing regularity on large domains of the parameter space.

We found that optimized shared weights yield highly applicable initial weights when considering nuclear geometries for which the wavefunction model was not previously optimized. Even for molecules such as ethene, pre-training with shared optimization makes it possible to reach an MRCI-F12(Q)/cc-pVQZ-F12 reference calculation up to chemical accuracy after only a few hundred optimization epochs. A possibly attractive route towards making VMC optimization tractable for more complex molecules could be to pre-train large parts of a wavefunction model on small, computationally cheap, systems. In our experiments, we found that the optimization of H$_{10}$ wavefunctions can in fact be improved significantly when using shared optimization for a set of H$_6$ geometries to pre-train the respective models. In the case of ethene, however, we could only see small improvements during early optimization when parts of the respective model were pre-trained on sets of smaller methane geometries. Due to the fact that any optimization of DNN-based wavefunction models is highly sensitive to changes in the architecture and optimization hyperparameters, we would consider our findings as preliminary evidence that merits further research towards the development of a kind of universal wavefunction, whose neural network parts were optimized for a great number of diverse molecular geometries and which is therefore capable of closely of approximating wavefunctions of the ground state for many physical systems after only a brief step of additional optimization.

Our results do not provide a conclusive answer whether for the investigated sets of molecular geometries, the considered weight-sharing setups actually limit the capability of our model to approximate the true wavefunctions of the ground state. In general, we would expect this to be the case, but judging from our experimental results, the loss of expressiveness introduced by weight-sharing regularization might often be negligible. This is evidenced by the fact that across all experiments, sharing 95\,\% of model weights yields the same or even lower energies than the respective independent optimizations. It is furthermore not clear yet, what an optimal algorithm that exploits weight-sharing for a given task could look like. Based on our results so far, for an exhaustive study of the PES of a molecule, we would suggest a procedure where -- possibly guided by estimates of the forces on the nuclei -- geometries of interest are iteratively included in a shared optimization, and which is eventually concluded by an additional step of independent optimization for some or all of the considered geometries. 

The proposed method of weight-sharing is not limited to the specific architecture used in this work but could potentially be exploited for any neural network-based wavefunction model. Due to the interesting regularization properties, it could even be beneficial to apply weight-sharing in a context where only the wavefunction for a single molecular geometry is of interest.

A potential drawback of the proposed method in practice is that shared optimization can not easily be parallelized across multiple devices (GPUs or CPUs), because each geometry is dependent on updates from all other geometries. One possibility to overcome this issue would be to consider an average loss across all geometries during gradient descent. Such a loss could easily be parallelized by using a separate device for each geometry. In our current implementation of the DeepErwin framework, however, for each epoch only a single geometry is considered during shared optimization, and it is therefore only possible to distribute the Monte Carlo samples within a batch across multiple devices, as it is common practice \cite{FermiNet2}.

\section{Methods}
\sloppy
For a molecule with $\nnuc$ nuclei, $\nsu$ spin-up electrons, and $\nsd$ spin-down electrons, we write $\bm{r} = (r_1,\ldots , r_{\nsu}, \ldots, r_{\nsu + \nsd})$ to denote the set of Cartesian electron coordinates, and $\bm{R} = (R_1,\ldots , R_{\nnuc})$ for the set of coordinates of nuclei. The electron coordinates $\bm{r}$ are always assumed to be ordered such that the first $\nsu$ entries correspond to spin-up electrons, while the last $\nsd$ entries are coordinates of spin-down electrons. 
We write $\nel = \nsu + \nsd$ for the total number of electrons and $Z_i$ for the charge of the $i$-th nucleus.
%
%
\subsection{Wavefunction model}
\label{subsec:architecture}
\begin{figure}[!tbp]
    \centering
    \includegraphics[width=1.0\columnwidth]{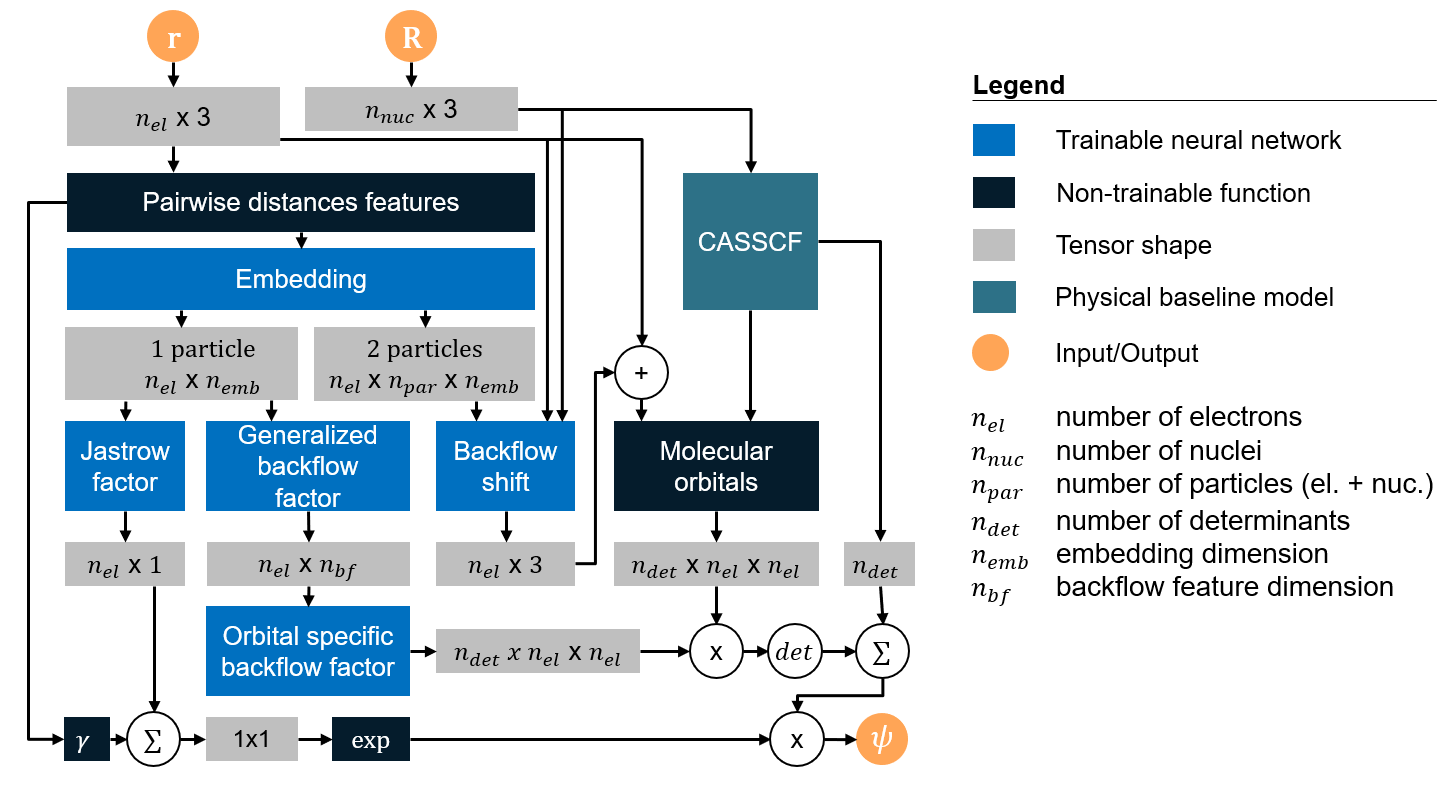}
    \caption{Overview of the computational architecture implemented in DeepErwin. Spin dependance has been omitted for clarity, but in practice there are two parallel streams for spin-up and spin-down respectively.}
    \label{fig:architecture}
\end{figure}

The model implemented in DeepErwin is closely related to the recently proposed PauliNet \cite{Hermann2020}. Let $\theta$ denote the set of all free (trainable) parameters in the model and $\ndet$ the number of enhanced Slater determinants. With a high-dimensional embedding of electron coordinates $\bm{x}(\bm{r}; \bm{R}) = (x_1, \ldots, x_{\nsu}, \ldots, x_{\nsu + \nsd})$ and an explicit term $\gamma(\bm{r})$ for cusp correction in the Jastrow factor, a realization $\psi_\theta$ of the DeepErwin wavefunction model can be written as
%
%
\begin{align}
\label{wavefunction}
    \psi_{\theta}(\bm{r}) = e^{J(\bm{x}(\bm{r}; \bm{R})) + \gamma(\bm{r})} \sum_{d = 1}^{\ndet} \alpha_d \det \bigl[ \bm{\Phi}^{\uparrow}_{d}\left(\bm{r}, \bm{x}(\bm{r}; \bm{R})\right) \bigr] \det \bigl[ \bm{\Phi}^{\downarrow}_{d}\left(\bm{r}, \bm{x}(\bm{r}; \bm{R})\right) \bigr],
\end{align}
where $\alpha_d \in \mathbb{R}$ is a trainable weight, the scalar function $J$ defining the Jastrow factor is represented by two fully connected feedforward neural networks, and Slater determinants for spin-up electrons are defined via
\begin{align}
\label{input_determinant}
\bm{\Phi}^{\uparrow}_{d}\left(\bm{r}, \bm{x}(\bm{r}; \bm{R})\right) =
\begin{bmatrix}
\varphi^{\uparrow, d}_{1} \bigl(r_1 + s_{1}(\bm{r}, \bm{x}; \bm{R}) \bigr) \eta_{1}^{\uparrow, d}(x_1) & \cdots & \varphi^{\uparrow, d}_{1} \bigl(r_{\nsu} + s_{\nsu}(\bm{r}, \bm{x}; \bm{R}) \bigr) \eta_{1}^{\uparrow, d}(x_{\nsu})\\
\vdots & \ddots & \vdots \\
\varphi^{\uparrow, d}_{\nsu} \bigl(r_1 + s_{1}(\bm{r}, \bm{x}; \bm{R}) \bigr) \eta_{\nsu}^{\uparrow, d}(x_1) & \cdots & \varphi^{\uparrow, d}_{\nsu} \bigl(r_{\nsu} + s_{\nsu}(\bm{r}, \bm{x}; \bm{R}) \bigr) \eta_{\nsu}^{\uparrow, d}(x_{\nsu})
\end{bmatrix}.
\end{align}
Matrices $\bm{\Phi}^{\downarrow}_{d}\left(\bm{r}, \bm{x}(\bm{r}; \bm{R})\right)$ for spin-down electrons can be defined analogously. The single electron orbitals $\varphi_i^{\uparrow, d}$ are obtained from the $\ndet$ most significant Slater determinants from a CASSCF method and remain fixed throughout optimization. The backflow shifts $s_{i}$ as well as the backflow factors $\eta_{i}^{\uparrow, d}$ are represented by fully connected feedforward neural networks. The embedded coordinate $x_i$ of the $i$-th electron takes into account all particle positions in the system independent of particle type and spin. An overview of the model given by eq.~\eqref{wavefunction} is shown in Figure~\ref{fig:architecture}. Further details regarding our implementation of the coordinate embedding, the Jastrow factor, backflow transformation, as well as cusp correction are given below.
%
%
%
%
%
\subsubsection{Electron coordinate embedding}
\label{subsubsec:embedding}
The embedding $\bm{x}(\bm{r}; \bm{R})$ is a slightly simplified version of the SchNet embedding used for PauliNet \cite{NIPS2017_303ed4c6, Hermann2020}. For brevity, we extend our notation to also include the nuclear coordinates in the coordinates $r_j$  by defining $r_j = R_{j-\nel}$ for $\nel < j \leq \nnuc$. To embed the coordinates $r_i$ of the $i$-th electron, we consider input features based on pairwise differences and distances with respect to all other particles in the system. Let $i \in \lbrace1, \dots, \nel\rbrace$, $j \in \lbrace1, \dots, \nel + \nnuc \rbrace$, and $\nrbf$ denote the number of radial basis features. We use $\vv{r}_{ij} = r_i - r_j$, and $r_{ij} = |\vv{r}_{ij}|$ to denote pairwise differences and distances, respectively, and define the pairwise feature vector
\begin{align}
\label{eq:pwfeatures}
    h_{ij} = \left(e^{-(r_{ij}-\mu_1)^2/\sigma_1^2},\ldots,e^{-(r_{ij}-\mu_{\nrbf})^2/\sigma_{\nrbf}^2}, \frac{1}{r_{ij} + 0.01} \right) \in \mathbb{R}^{\nrbf + 1 + 3},
\end{align}
where the mean and variance parameters are defined as
\begin{equation}
    \mu_k = c q_k^2, \ \text{and} \  \sigma_k = \frac{1}{7}(1 + cq_k),
\end{equation}
respectively, for an index $k \in \lbrace1, \dots, \nrbf\rbrace$, a parameter $q_k$ that is chosen from an equidistant grid of the interval $[0, 1]$, and a cutoff parameter $c \in \mathbb{R}$.

For a fixed integer $L$ and an embedding dimension $\nemb$, let $\left(g^l, w^l_{\textrm{same}}, w^l_{\textrm{op}}, w^l_{\textrm{nuc}}\right)_{l=0}^L$ and $\left(f^l_{\textrm{same}}, f^l_{\textrm{op}}\right)_{l=1}^L$ denote sequences of vector-valued functions, where each function is represented by a fully connected feedforward neural network with output dimension $\nemb$. To embed the coordinates of the $i$-th electron based on the pairwise feature vectors $h_{ij}$, we use $\odot$ to denote element-wise multiplication and define for $i \in  \{1, \ldots, \nel\}$
\begin{align}
    x_i^0 &= g^0 \left(\sum_{\substack{j=1\\j\neq i}}^{\nel} w^{0}_{\sigma_{ij}} (h_{ij}) \odot f^{0}_{\sigma_{ij}}  +\sum_{j = \nel + 1}^{{\nel + \nnuc}} w^{0}_{\textrm{nuc}} (h_{ij}) \odot f^{0}_{Z_{j - \nel}}\right), \\
    x_i^{l} &= g^l \left(\sum_{\substack{j=1\\j\neq i}}^{\nel} w^{l}_{\sigma_{ij}} (h_{ij})\odot f^{l}_{\sigma_{ij}}(x_j^{l-1}) + \sum_{j = \nel + 1}^{{\nel + \nnuc}} w^{l}_{\textrm{nuc}} (h_{ij}) \odot f^{l}_{Z_{j - \nel}} \right) \ \text{with} \ 1 \leq l \leq L, \label{eq:embl}
\end{align}
where $\sigma_{ij} = \textrm{'same'}$ for same-spin pairs of electrons and $\sigma_{ij} = \textrm{'op'}$ for pairs of electrons with opposite spin, and where $f^0_{\textrm{same}}$, $f^0_{\textrm{op}}$, and $f^{0}_{Z_{j - \nel}}, \ldots, f^{L}_{Z_{j - \nel}}$ denote trainable vectors of length $\nemb$. The embedding of the $i$-th electron coordinates $r_i$ is eventually defined as
\begin{equation}
    x_i = x^L_{i}.
\end{equation}
The embedding originally applied in PauliNet has an additional residual term in eq.~\eqref{eq:embl} and considers specific functions $\left(g^l\right)_{l = 0}^L$ for same-spin, opposite-spin, and  nuclear input channels.
\subsubsection{Backflow transformation}
Based on the embedding $\bm{x}(\bm{r}; \bm{R})$, our model applies spin-dependent backflow shifts and factors to the single electron orbitals in the Slater determinants (cf. eq.~\eqref{input_determinant}). For simplicity, this section only considers the spin-up case.

Let $\eta^\uparrow_{\textrm{gen}}$ denote a vector-valued function that is represented by a fully connected feedforward neural network, and $\omega_{\textrm{bf}} \in \mathbb{R}$ a single spin-independent trainable weight. For the $i$-th orbital and the $j$-th electron in the $d$-th Slater determinant, the backflow factor is computed as
\begin{equation}
\eta_{i}^{\uparrow, d}(x_j) = 1 + \omega_{\textrm{bf}}   \widehat{\eta}^{\uparrow, d}_{i}\bigl(\eta^\uparrow_{\textrm{gen}}(x_j)\bigr),
\end{equation}
where $\widehat{\eta}^{\uparrow, d}_{i}$ denotes an orbital-specific function that is also represented by a feedforward neural network.

The inner function $\eta^\uparrow_{\textrm{gen}}$ can be seen as an extension of the embedding layer. Apart from electron spin orientation, it remains unchanged across electrons and determinants and is hence called \emph{general backflow factor}. The outer function $\widehat{\eta}^{\uparrow, d}_{i}$, on the other hand, is specific to the $i$-th orbital in the $d$-th determinant obtained from a CASSCF method. The motivation behind defining the backflow factor via the composition of these two functions was to maximize the number of neural networks weights than can possibly be shared across nuclear geometries. Note that for two distinct nuclear geometries, CASSCF does in general not yield the same number of unique orbitals. While this does not raise an issue for the general backflow factor $\eta^\uparrow_{\textrm{gen}}$, it implies that sharing the neural network weights that define $\widehat{\eta}^{\uparrow, d}_{i}$ across nuclear geometries is in general not possible in a meaningful way. 

Based on ideas from \cite{rios2006inhomogeneous}, the backflow shift is based on rotation-invariant features and rotation-equivariant pairwise differences. It is furthermore split into an electron-electron and an electron-nucleus part. Besides the embedding, we additionally use side products of the embedding network as inputs and define the following feature vectors
\begin{align}
    f^{\textrm{el}}_{i} = \bigl (w^{L}_{\sigma_{i1}} (h_{i1})\odot f^{L}_{\sigma_{i1}}(x_1^{L-1}), \dots, w^{L}_{\sigma_{i\nel}} (h_{i\nel})\odot f^{L}_{\sigma_{i\nel}}(x_{\nel}^{L-1}) \bigr) 
\end{align}
and 
\begin{align}
    f^{\textrm{nuc}}_{i} = \bigl (w^{L}_{\textrm{nuc}} (h_{i (\nel +1)}) \odot f^{L}_{Z_{1}}, \dots, w^{L}_{\textrm{nuc}} (h_{i (\nel + \nnuc)}) \odot f^{L}_{Z_{\nnuc}} \bigr)
\end{align}
for $ i \in \lbrace1, \dots, \nel\rbrace$. All features are obtained without additional cost at the end of the embedding loop (cf. \ref{eq:embl}). The electronic part of the shift for the $i$-th electron is defined as 
\begin{equation}
    s^{\textrm{el}}_i(\bm{r}, x_i, f^{\textrm{el}}_{i}) = \sum_{\substack{k=1\\k\neq i}}^{\nel} \widehat{s}_{\textrm{el}} \bigl(x_i, f^{\textrm{el}}_{i} \bigr ) \frac{\vv{r}_{ik}}{ 1 + r_{ik}^3}.
\end{equation}
In contrast to the backflow factor, we do not differentiate between different spins to reduce complexity. Similarly, the electron-nuclear shift is computed via
\begin{equation}
    s^{\textrm{nuc}}_i(\bm{r}, \bm{R}, x_i, f^{\textrm{nuc}}_{i}) = \sum_{k = \nel + 1}^{\nnuc + \nel} \widehat{s}_{\textrm{nuc}} \bigl(x_i, f^{\textrm{ion}}_{i} \bigr) \frac{\vv{r}_{ik}}{ 1 + r_{ik}^3},
\end{equation}
where $\widehat{s}_{\textrm{el}}$ and $\widehat{s}_{\textrm{nuc}}$ are represented by feed-forward neural networks. A decay in the vicinity of nuclei ensures that the applied backflow shifts do not lead to a violation of the Kato cusp condition \cite{KatoCuspCondition}. The complete backflow shift is computed by
\begin{equation}
    s_i(\bm{r}, \bm{R}, x_i, f^{\textrm{el}}_{i}, f^{\textrm{nuc}}_{i}) = \omega_{\textrm{sh}} \prod_n \tanh{ \left(2Z_n|r_i - R_n| \right)^2} \bigl (s^{\textrm{el}}_i(x_i, f^{\textrm{el}}_{i}) +  s^{\textrm{nuc}}_i(x_i, f^{\textrm{nuc}}_{i}) \bigr ),
\end{equation}
with a spin-independent trainable weight $\omega_{\textrm{sh}} \in \mathbb{R}$.
\subsubsection{Jastrow factor}
With two spin-dependent scalar functions $J^{\uparrow}$ and $J^{\downarrow}$, that are both represented by fully connected feedforward neural networks, the Jastrow factor (cf. eq.~\eqref{wavefunction}) is defined as
\begin{equation}
    J\left(\bm{x}(\bm{r}; \bm{R}) \right) = \sum_{i = 1}^{\nsu} J^{\uparrow}(x_i) + \sum_{i = \nsu + 1}^{\nel} J^{\downarrow}(x_i).
\end{equation}
\subsubsection{Cusp correction}
 The Kato cusp condition \cite{KatoCuspCondition} is a necessary condition for eigenstates of the Hamiltonian $H$. It ensures that the local energies of a wavefunction $\psi$ are finite by forcing the kinetic energy term $\nabla^2 \psi$ to diverge in such a way that it exactly cancels the divergence caused by the potential energy term $Z_n / |r_i - R_n|$ when the $i$-th electron approaches the $n$-th nucleus. The orbitals $\varphi^{\uparrow, d}_{i}$, respectively $\varphi^{\downarrow, d}_{i}$, yielded by a CASSCF method (cf. eq.~\ref{input_determinant}), do in general not lead to enhanced Slater determinants that satisfy the cusp condition. To address this issue, we follow the approach outlined in \cite{Ma2005_CuspScheme}: Within a radius $R_\text{cusp}$ around the nuclei, we replace the molecular orbitals by an exponentially decaying function which satisfies the Kato cusp condition, transitions smoothly into the orbitals at $R_\text{cusp}$, and minimizes the variance of the local energy of this orbital. Cusp correction for the enhanced Slater determinants is performed after the initial set of orbitals has been obtained from a CASSCF calculation and remains fixed during the optimization of the wavefunction parameters $\theta$.

To account for electron-electron cusps in the Jastrow factor (cf. eq.~\ref{wavefunction}), we use an explicit term
\begin{equation}
    \gamma(\bm{r}) = \sum_{i=1}^{\nel} \sum_{j=i+1}^{\nel}\frac{r_{ij}}{r_{ij} + 1},
\end{equation} 
similar to the one applied in \cite{Han_2019}. 
\subsection{Variational Monte Carlo}
We use a standard variational Monte Carlo approach to optimize our wavefunction ansatz. Let
\begin{equation}
    H = E_\textrm{kin} + E_{\textrm{pot}}
\end{equation}
denote the electronic Hamiltonian as obtained within the Born-Oppenheimer approximation, with
\begin{align}
&E_\textrm{kin} = -\frac{1}{2}\sum_{i} \nabla^2_{r_i},\\
    &E_{\textrm{pot}} = \sum_{i>j} \frac{1}{|{r_i - r_j}|} + \sum_{n>m}\frac{Z_n Z_m}{|{R_n - R_m}|} - \sum_{i,n} \frac{Z_n}{|{r_i - R_n}|},
\end{align}
where $E_\textrm{kin}$ accounts for the kinetic energy of the electrons, and $E_{\textrm{pot}}$ accounts for the attraction and repulsion between particles in the system. By the Rayleigh-Ritz variational principle, it holds for any wavefunction $\psi_\theta$ that its energy $E_\theta$ is greater than or equal to the ground state energy $E_0$ of the eigenfunction of $H$ associated to the smallest eigenvalue, that is
\begin{equation}
\label{ground state_energy}
    E_\theta = \int \frac{\psi_\theta(\bm{r}) H \psi_\theta(\bm{r})}{\Omega_\theta} d\bm{r} \geq E_0,
\end{equation}
where $\Omega_\theta = \int \psi_\theta(\bm{r})^2 d\bm{r}$ denotes a normalization factor.

To evaluate the energy $E_\theta$ for a given set of parameters $\theta$, we use Markov chain Monte Carlo integration (MCMC) and sample electron coordinates according to the probability density 
\begin{equation}
    \label{eq:ptheta}
    p_\theta(\bm{r}) = \frac{\psi_\theta(\bm{r})^2}{\Omega_\theta}.
\end{equation}
This allows us to express the total energy $E_\theta$ as the expected value of a local energy
\begin{equation}
\label{eq:eloc}
E_{\textrm{loc}}(\bm{r}) = \frac{H \psi_\theta(\bm{r})}{\psi_\theta(\bm{r})},
\end{equation}
in the sense that
\begin{equation}
    \label{eq:energyfromlocal}
    E_\theta = \int E_{\textrm{loc}}(\bm{r}) p_\theta(\bm{r}) d \bm{r} = \Braket{E_\text{loc}} \approx \frac{1}{N}\sum_{k=1}^N E_\textrm{loc}(\bm{r}^k),
\end{equation}
where $N$ denotes the number of sampled electron coordinates and $\bm{r}^k \sim p_\theta$.

To optimize the parameters $\theta$, we use K-FAC \cite{martens2015optimizing} which was already implemented in FermiNet \cite{FermiNet} to minimize the local energy $E_\textrm{loc}(\bm{r}^k)$ at the location of electron coordinates $\bm{r}^k$. DeepErwin also supports the limited-memory Broyden–Fletcher–Goldfarb–Shanno algorithm (L-BFGS \cite{LBFGS}), which is another second-order method, as well as standard first-order stochastic gradient descent (SGD) as implemented by the ADAM algorithm \cite{Adam}. After each optmization epoch, that is, after each sample of coordinates was used exactly once for a gradient descent step, the coordinates $\bm{r}^k$ are resampled to reflect the updated probability distribution $p_\theta$. By applying this procedure iteratively for a large number of epochs, we eventually obtain parameters $\theta$ such that the wavefunction $\psi_\theta$ closely approximates the wavefunction of the ground state for the considered molecule.

To increase numerical stability, the implementation of our ansatz does not directly model $\psi_\theta$, but the logarithm of its square. The local energy can then be computed as
\begin{equation}
     E_\textrm{loc}(\bm{r})=  E_{\textrm{pot}}(\bm{r}) - \frac{1}{4}\nabla_{\bm{r}}^2 \phi_\theta(\bm{r}) - \frac{1}{8} \left(\nabla_{\bm{r}} \phi_\theta\right)^2(\bm{r}),
\end{equation}
where $\phi_\theta = \log(\psi_\theta^2)$.

Although the local energy already contains second derivatives (with respect to $\bm{r}$), calculating the gradient with respect to $\theta$ does not require the calculation of third derivatives, because $H$ is Hermitian \cite{EnergyGradient_1976}. Precisely, it holds that
\begin{align}
    \nabla_\theta E_\theta &= \Braket{E_\text{loc}  \nabla_\theta \phi_\theta} - \Braket{E_\text{loc}} \Braket{ \nabla_\theta \phi_\theta}\\
    &\approx \frac{1}{N} \sum_{k=1}^N E_\text{loc}(\bm{r}^k) \nabla_\theta \phi_\theta(\bm{r}^k) - \frac{1}{N^2}\left(\sum_{k=1}^N E_\text{loc}(\bm{r}^k) \right) \left(\sum_{k=1}^N \nabla_\theta \phi_\theta(\bm{r}^k) \right),
\end{align}
for $N$ samples of electron coordinates $\bm{r}^k \sim p_\theta$.

To approximate the distribution of electron coordinates defined by the density $p_\theta$, we typically use about 2k independent MCMC chains (walkers), where each walker is initialized before optimization with a large number ($\sim$ 1k) of burn-in steps. During optimization, walkers are updated after every epoch with a small number ($\sim$ 10) of additional MCMC steps. To precisely estimate the energy $E_\theta$ after optimization has concluded, we apply a similar procedure, but collect the local energies for all walkers from roughly 1k intermediate steps in the respective MCMC chain.

For single steps in the MCMC chains, we use a Metropolis Hastings algorithm \cite{Hastings1970}. Given a current walker state $\bm{r}$, a proposal state $\bmrprop$ for the Metropolis Hastings algorithm is generated according to the probability density
\begin{equation}
p(\bmrprop | \bm{r}) = \prod_{i=1}^{\nel}p_{\mathcal{N}}\left(\rprop_i | \mu = {r}_i, \sigma^2 = (d_i)^2\right),
\end{equation}
where $p_{\mathcal{N}}$ denotes the density of a three-dimensional normal distribution and the variance parameters $d_i$ are defined for fixed parameters $d_0, d_{\textrm{min}}, d_{\textrm{max}}, \delta \geq 0$ as
\begin{equation}
\label{eq:mhvariance}
    d_i = \delta \min \left(d_{\textrm{min}} + \frac{|{r_i - R_{n}}|}{d_0}, d_{\textrm{max}}\right),
\end{equation}
where $R_n$ denotes the position of the nucleus closest to $r_i$, that is, $n = \argmin_{j=1,\dots , \nnuc}|{r_i - R_{j}}|$. We then consider the canonical acceptance probability
\begin{equation}
p_{\textrm{acc}}(\bmrprop | \bm{r}) = \min\left\{1, \frac{p(\bm{r} | \bmrprop )p_\theta(\bmrprop)}{p(\bmrprop | \bm{r})p_\theta(\bm{r})}\right\},
\end{equation}
and for a sample $\alpha \in [0,1]$ from a uniform distribution, the proposed sample $\bmrprop$ is accepted in the case $p_{\textrm{acc}}(\bmrprop | \bm{r}) > \alpha$ and rejected otherwise.

The parameters $d_i$ defined in eq.~\eqref{eq:mhvariance} can be seen as step size parameters that regulate the average distance between electron coordinates $\bm{r}$ and a proposal $\bmrprop$. In general, smaller step sizes lead to higher acceptance rates for proposed samples. The parameter $\delta$ is a general step size parameter that is gradually adapted during optimization to yield an average acceptance rate of 50\%. The second factor in eq.~\eqref{eq:mhvariance} was specifically designed to take into account that wavefunctions are usually most complex in the proximity of a nucleus. That is, the step size $d_i$ is chosen for each electron $i$ to depend on its distance to the closest ion to encourage smaller steps close to the nuclei, where the wavefunction varies rapidly, and larger steps, when an electron is further away from the nuclei. 

Note that the step sizes $d_i$, and thus the proposal probability, depend on the electron positions $\bm{r}$. In general it is therefore not true that $p(\bmrprop | \bm{r}) = p(\bm{r} | \bmrprop)$. However, due to our choice for the acceptance probability $p_{\textrm{acc}}$, the so-called detailed balance condition is still satisfied. That is, being in a state $\bm{r}$ and transitioning to $\bmrprop$ is as probable as being in $\bmrprop$ and transitioning to $\bm{r}$.
%
%
%
%
\subsection{Forces}
For a given molecule, let $\psi_0$ denote the wavefunction of the ground state, that is, $\psi_0$ is the eigenfunction of the Hamiltonian $H$ associated to the smallest eigenvalue. To calculate the electronic forces acting on the $m$-th nuclei, we apply the Hellmann-Feynman theorem \cite{HellmanFeynman} and compute
\begin{align}
    F_m &= -\nabla_{R_m} E_0 = - \frac{1}{\Omega_0}\int \psi_0(\bm{r}) \left((\nabla_{R_m} H\right) \psi_0)(\bm{r}) d\bm{r}\\
    &= Z_m \frac{1}{\Omega_0} \int \psi_0(\bm{r})^2 \left(\sum_i \frac{r_i-R_m}{|r_i - R_m|^3} \right) d\bm{r} - \sum_{n \neq m} Z_n \frac{R_n - R_m}{|R_n - R_m|^3}\\
    &\approx Z_m \left(\frac{1}{N} \sum_{k} \sum_i \frac{r^k_i-R_m}{|r^k_i - R_m|^3} -\sum_{n \neq m} Z_n \frac{R_n - R_m}{|R_n - R_m|^3} \right), \label{eq:forces}
\end{align}
for $N$ samples of electron coordinates $\bm{r}^k \sim \psi_0(\bm{r})^2/\Omega_0$, and where $\Omega_0 = \int \psi_0(\bm{r})^2 d\bm{r}$ denotes the $L^2$-norm of $\psi_0$. 

Since it is no longer necessary to compute derivatives of the wavefunction, evaluating eq.~\eqref{eq:forces} should be relatively easy. However, unlike the local energy (cf. eq.~\eqref{eq:energyfromlocal}), which has zero variance for eigenstates of the Hamiltonian, naive Monte Carlo sampling can not be applied in this case due to the divergence of $|r_i - R_k|^{-3}$ when the $i$-th electron approaches the $k$-th nucleus.

This issue can be addressed by observing that for each diverging term on one side of a nucleus, there is an equally diverging term with the opposite sign on the other side of the nucleus. Different variance reduction methods have been proposed to exploit this property, such as fitting the force density close to a nucleus with a function that is constrained to be zero at $r_i = R_k$ \cite{Chiesa2005_PRL_force_fit}, or antithetic sampling \cite{Chiesa2005_PRL_force_fit, Kalos_MonteCarloMethodsBook_antithetic}. 

We minimize the variance of force samples by combining antithetic sampling with a truncated $1/r$ potential: For each sample of electron coordinates $r^k$ yielded by a Markov chain, we also consider an additional sample $\widehat{\bm{r}}^{k}$ in which each electron within a distance of $R_\text{core}$ to the closest nucleus is mirrored to the opposite side of this nucleus.
\begin{equation}
    \widehat{r}^{k, i} = r^k_i + 2 (R_m - r^k_i)
\end{equation}
for $i \in \{1,\ldots,\nel\}$, and $m = \argmin_n |r_i - R_n|$. Therefore, whenever an electron comes close to a nucleus and thus generates a large force in one direction, there will always be a cancelling contribution in the opposite direction. Note that the samples $\widehat{\bm{r}}^{k, i}$ do not necessarily have the same probability as the original sample $\bm{r}^k$. Their contribution to the Monte Carlo estimate is thus weighted by $\frac{\psi(\widehat{\bm{r}}^{k, i})^2}{\psi(\bm{r}^k)^2}$.
Additionally we avoid numerical instabilities by replacing the raw Coulomb forces with a scaled version that decays towards zero, as electrons approach a nucleus:
\begin{equation}
    \frac{r^k_i-R_m}{|r^k_i - R_m|^3} \longrightarrow \frac{r^k_i-R_m}{|r^k_i - R_m|^3} \tanh^3 \left(\frac{|r^k_i - R_m|}{R_\textrm{cut-off}}\right)
\end{equation}
\subsection{Shared optimization of model parameters}
For $N$ distinct nuclear geometries $\bm{R}^1, \bm{R}^2, \dots, \bm{R}^{N}$, let $\thetash$ denote the set of model parameters that are shared across all geometries, and $\widehat{\theta}^k$ denote the set of parameters that are specific to the $k$-th set of nuclear coordinates $\bm{R}^k$. The full set of parameters for the $k$-th geometry can then be written as $\theta^k = (\thetash, \widehat{\theta}^k)$, and the associated realization of the wavefunction model is denoted by $\psi_{\theta^k}$. During each shared optimization epoch, we consider a single nuclear geometry $\bm{R}^k$ and update the respective model weights $\theta^k$ with respect to the local energies of the wavefunction $\psi_{\theta^k}$ at MCMC walker positions that were sampled from the probability density $p_{\theta^k}$ (cf. eq.~\eqref{eq:ptheta}). That is, during each shared optimization epoch, not only the geometry-specific weights are updated, but also the shared weights $\thetash$, and therefore the wavefunctions for all nuclear geometries.




A straightforward way of deciding which geometry to consider for a shared optimization epoch is to employ a simple round-robin scheme. Note that for each eigenstate of the Hamiltonian, the local energies defined in eq.~\eqref{eq:eloc} have zero variance. Therefore, another approach to select geometries for a shared optimization epoch is to use the standard deviation of local energies as a proxy of how closely a realization of the wavefunction model already approximates the wavefunction of the ground state. By selecting the geometry for which the current wavefunction has the highest standard deviation in the local energies, we ensure that geometries for which the parameters have not yet been well optimized get more attention during optimization. In practice, we train a few initial epochs using the round-robin scheme to obtain a reliable starting point for each wavefunction and then switch to the standard-deviation-based scheme, to ensure homogeneous convergence of the accuracy across all geometries.

In our implementation of the shared optimization scheme, we use a single instance of the optimizer to update the set of shared weights $\thetash$, as well as an additional instance for each of the geometry-specific sets of model parameters. In particular, due to the fact that geometry-specific weights receive significantly less gradient descent updates than shared weights, the learning rate for the geometry-specific optimizers are usually chosen about 10 times larger than the learning rate for the optimizer of the shared weights (cf. Table~\ref{table:hyperparams}).


After the shared optimization process, the wavefunctions can again be treated as fully independent realizations of our wavefunction model and are not constrained by any additional dependencies. In particular, this means that the evaluation of their exact energy can easily be parallelized across geometries.
\subsection{Reference calculations}
\label{subsec:molpro_ref_calcs}
In order to validate our deep learning method, we compared the obtained energies to reference values, which we computed using the MOLPRO package \cite{MOLPRO-WIREs,MOLPRO}. We employed both single- and multi-reference explicitly correlated F12 methods: Coupled cluster with singles, doubles and perturbative triples (CCSD(T)-F12) \cite{Adler_et_al}
and multi-reference configuration interaction (MRCI-F12) \cite{Shiozaki_et_al}. 
The basis set cc-pVQZ-F12 \cite{Peterson_et_al} 
was used for all geometries of H$_4^+$, H$_6$, and C$_2$H$_4$. Due to convergence issues with this basis set for some geometries of H$_{10}$, the smaller cc-pVTZ-F12 basis set was employed for this molecular system. 
For the MRCI-F12 calculations, a CASSCF reference was used with a full-valence active space (H$_4^+$: 3 electrons in 3 orbitals abbreviated as (3, 3), H$_6$: (6,6), H$_{10}$: (10,10)). A full-valence active space was prohibitively expensive for C$_2$H$_4$, which is why we resorted to an active space of (2,2). The recommended GEM\_BETA coefficients were used (i.e., for MRCI-F12(Q)/cc-pVQZ-F12 calculations GEM\_BETA = 1.5 a$_0^{-1}$ and for MRCI-F12(Q)/cc-pVTZ-F12 calculations GEM\_BETA = 1.4 a$_0^{-1}$) \cite{Hill_et_al}. 
The Davidson-corrected \cite{DavidsonCorrection} 
energy values (MRCI-F12(Q)) were extracted, as provided by the energyd variable in MOLPRO.

We note that these accurate methods and basis sets were only used to validate our results, but not as a starting point for our deep learning method. The orbitals used as a starting point for our method were generated using the CASSCF method implemented in PySCF \cite{PySCF}, using a 6-311G Pople basis set.

\subsection{Computational settings for DeepErwin}
Reference values for the results shown in Fig.~\ref{fig:parallel_speedup} were obtained via MRCI-F12(Q)/cc-pVQZ-F12. See Sec.~\ref{subsec:molpro_ref_calcs} for more details. 
    
In all cases, we used as reference the described method from section \ref{subsec:molpro_ref_calcs}. However, for the initialization from the H$_{10}$ run two geometries were excluded from the reference set, since these were also included in the pre-trained geometry grid.

The main hyperparameters used for all computations are listed in Table~\ref{table:hyperparams}. Detailed counts of total, trainable, and shared parameters for the used wavefunction models for the four main molecules considered in our numerical experiments are compiled in Table~\ref{table:parameternumbers}.

The DeepErwin package alongside a detailed documentation is available on the Python Package Index (PyPI) and github (\url{https://github.com/mipunivie/deeperwin}).
\renewcommand*{\arraystretch}{1.2}
\begin{table}
\caption{Hyperparameter settings for DeepErwin.\label{table:hyperparams}}

\begin{tabularx}{\textwidth}{|X|X|c|}
\hline
\multirow{2}{*}{ \textbf{Baseline method (CASSCF)}}
		 &\#\,determinants & 20\\ \cline{2-3}
		 &Basis set & 6-311G\\ \cline{2-3}
\hline
\hline
\multirow{7}{*}{ \textbf{Embedding}}
		 &\#\,radial basis features $\nrbf$   & 16\\ \cline{2-3}
		 &\#\,hidden layers $g$		                  & 1\\ \cline{2-3}
		 &\#\,hidden layers $w$		                  & 2\\ \cline{2-3}
		 &\#\,hidden layers $f$		                  & 2\\ \cline{2-3}
		 &\#\,neurons per layer $f, w, g$		      & 40\\ \cline{2-3}
		 &\#\,iterations $L$		                  & 2\\ \cline{2-3}
		 &Embed
		 ding dimension $x$			              & 64\\ \cline{2-3}
\hline
\hline
\multirow{4}{*}{ \textbf{Backflow / Jastrow}}
		 &\#\,hidden layers $\eta_{\text{gen}}, s, J$		                & 2\\ \cline{2-3}
		 &\#\,neurons per layer $\eta_{\text{gen}}, s, J$		            & $40$\\ \cline{2-3}
		 &\#\,hidden layers $\eta_{i}$		                & 0\\ \cline{2-3}
		 &\#\,neurons per layer $\eta_{i}$		            & $1$ \\ \cline{2-3}
\hline
\hline
\multirow{3}{*}{ \textbf{MCMC}}
		 &\#\,walkers                        & $2048$	\\ \cline{2-3}
		 &\#\,decorrelation steps            & 5	\\ \cline{2-3}
		 &Target acceptance probability         & 50\%	\\ \cline{2-3}
\hline
\hline
\multirow{4}{*}{ \textbf{Optimization general}}
		 &Optimizer		                        & KFAC\\ \cline{2-3}
		 &Damping 	                            & $5 \times 10^{-4}$ 	\\ \cline{2-3}
		 &Norm constraint	                    & $1 \times 10^{-3}$ 	\\ \cline{2-3}

		 &Batch size 	                        & $512$	\\ \cline{2-3}
\hline
\hline
\multirow{2}{*}{ \textbf{Independent optimization}} 		 &Initial learning rate $\text{lr}_0$	& $2 \times 10^{-3}$	\\ \cline{2-3}
		 &Learning rate decay   	            & $\text{lr}(t) = \frac{\text{lr}_0}{1+t/1000}$ 	\\ \cline{2-3}
\cline{2-3}
\hline
\hline
\multirow{4}{*}{ \shortstack[l]{\textbf{Shared optimization}\\  \textbf{(Sec.~\ref{subsec:sharedopt})}}}
		 &$\text{lr}_0$ for non-shared weights 	                    & $5 \times 10^{-3}$ 	\\ \cline{2-3}
		 &$\text{lr}_0$ for shared weights	                & $5 \times 10^{-4}$	\\ \cline{2-3}
		 &Learning rate decay	                &  $\text{lr}(t) = \frac{\text{lr}_0}{1+t/10000}$	\\ \cline{2-3}
		 &Training scheduler	                & Std. deviation	\\ \cline{2-3}
\hline
\hline
\multirow{3}{*}{\shortstack[l]{\textbf{Optimization after pre-training}\\ \textbf{(Sec.~\ref{subsec:reuse})}}}
		 &$\text{lr}_0$ for non-shared pre-trained weights	                    & $2\times 10^{-3}$ 	\\ \cline{2-3}
		 &$\text{lr}_0$ for shared pre-trained weights	                & $2\times 10^{-4}$	\\ \cline{2-3}
		 &Learning rate decay	                &  $\text{lr}(t) = \frac{\text{lr}_0}{1+t/1000}$	\\ \cline{2-3}
\hline
\hline
\multirow{2}{*}{\textbf{Force evaluation}} &Antithetic sampling radius $R_\text{core}$	& $0.2$	\\ \cline{2-3}
		 & Coulomb cutoff $R_\text{cut-off}$   	            & $0.01$ 	\\ \cline{2-3}
\cline{2-3}
\hline
\end{tabularx}
\end{table}
\begin{table}
\caption{Counts of total, trainable and shared model parameters for each of the four main molecules considered in the numerical experiments.\label{table:parameternumbers}}
\begin{tabularx}{\textwidth}{|X|l|r|r|r|r|}
\hline
Molecule&Optimization&\#\,total & \#\,non-trainable& \#\,trainable & \#\,shared\\
\hline
\multirow{3}{*}{ \textbf{H$_4^+$}}
& independent & \numprint{107548} & \numprint{368} (0.3\,\%) & \numprint{107180} & \numprint{0} (0.0\,\%)\\\cline{2-6}
& 75\,\% shared & \numprint{107548} & \numprint{368} (0.3\,\%) & \numprint{107180} & \numprint{83673} (77.8\,\%)\\\cline{2-6}
& 95\,\% shared & \numprint{107548} & \numprint{368} (0.3\,\%) & \numprint{107180} & \numprint{105920} (98.5\,\%)\\\cline{2-6}
\hline\hline
\multirow{3}{*}{ \textbf{H$_6$}}
& independent & \numprint{109228} & \numprint{788} (0.7\,\%) & \numprint{108440} & \numprint{0} (0.0\,\%)\\\cline{2-6}
& 75\,\% shared & \numprint{109228} & \numprint{788} (0.7\,\%) & \numprint{108440} & \numprint{83673} (76.6\,\%)\\\cline{2-6}
& 95\,\% shared & \numprint{109228} & \numprint{788} (0.7\,\%) & \numprint{108440} & \numprint{105920} (97.0\,\%)\\\cline{2-6}
\hline\hline
\multirow{3}{*}{ \textbf{H$_{10}$}}
& independent & \numprint{112140} & \numprint{2020} (1.8\,\%) & \numprint{110120} & \numprint{0} (0.0\,\%)\\\cline{2-6}
& 75\,\% shared & \numprint{112140} & \numprint{2020} (1.8\,\%) & \numprint{110120} & \numprint{83673} (74.6\,\%)\\\cline{2-6}
& 95\,\% shared & \numprint{112140} & \numprint{2020} (1.8\,\%) & \numprint{110120} & \numprint{105920} (94.5\,\%)\\\cline{2-6}
\hline\hline
\multirow{3}{*}{ \textbf{Ethene}}
& independent & \numprint{115868} & \numprint{3228} (2.8\,\%) & \numprint{112640} & \numprint{0} (0.0\,\%)\\\cline{2-6}
& 75\,\% shared & \numprint{115868} & \numprint{3228} (2.8\,\%) & \numprint{112640} & \numprint{83673} (72.2\,\%)\\\cline{2-6}
& 95\,\% shared & \numprint{115868} & \numprint{3228} (2.8\,\%) & \numprint{112640} & \numprint{105920} (91.4\,\%)\\\cline{2-6}
\hline
\end{tabularx}
\label{table:hyperparameters}
\end{table}
\section{Acknowledgements}
L.G. gratefully acknowledges support from the Austrian Science Fund (FWF I 3403) and the WWTF (ICT19-041). R.R. gratefully acknowledges support from the Austrian Science Fund (FWF M 2528). The computational results presented have been achieved using the Vienna Scientific Cluster (VSC).
%
%
\printbibliography
\end{document}


\section*{Supplementary information}
\subsection{Weight-sharing and pre-training with ADAM}
\begin{figure}[h]
    \centering
    \includegraphics[width=.95\columnwidth]{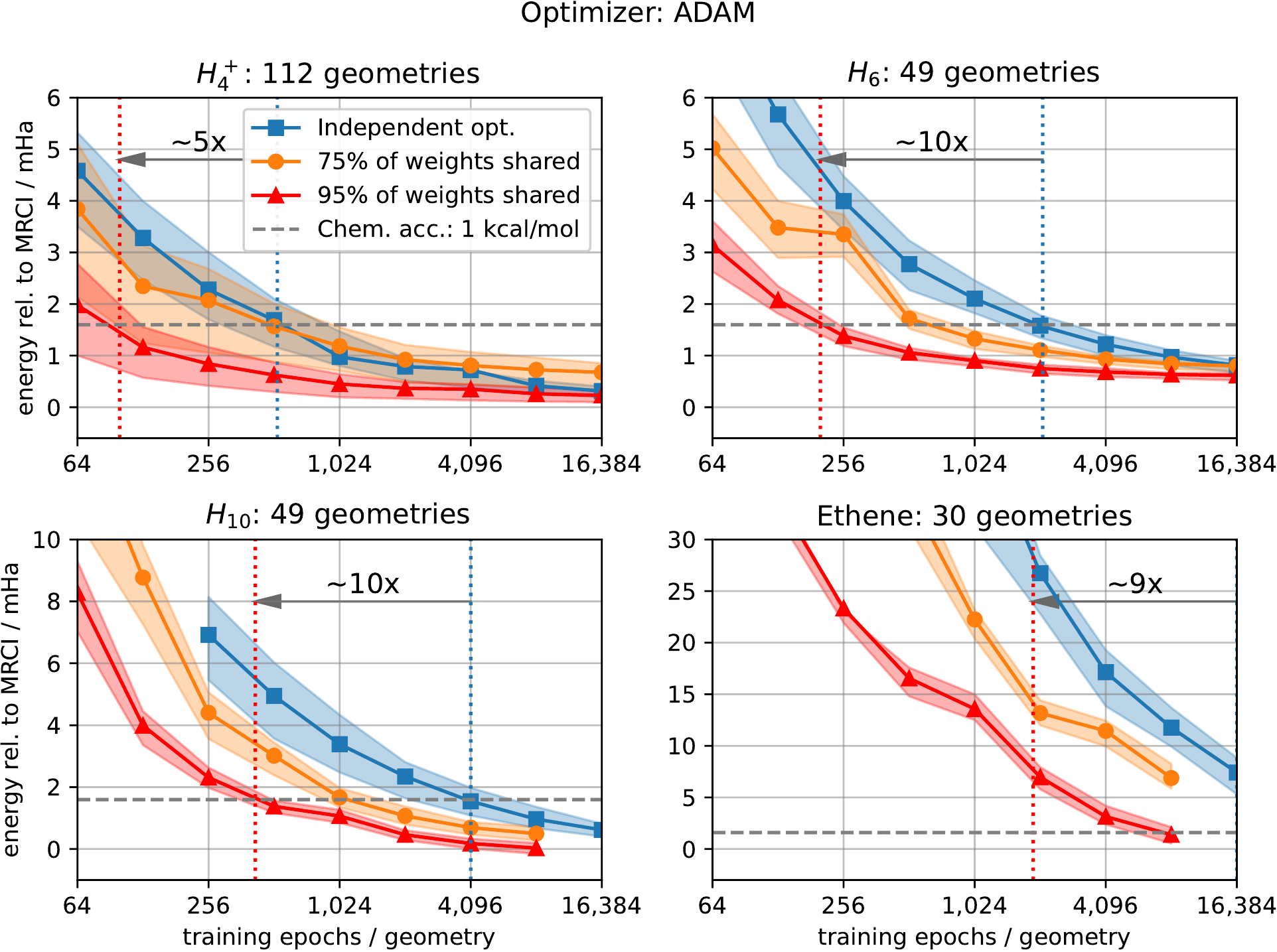}
    \caption{Mean evaluation error relative to the reference calculation (MRCI-F12(Q)/cc-pVQZ-F12) as a function of training epochs per geometry for four different sets of nuclear geometries. Shadings indicate the area between the 25\,\% and the 75\,\% percentile of considered geometries. For each method, we plot intermediary and final results for optimizations that ran for a total number of \numprint{16384} epochs per geometry.}
    \label{fig:parallel_speedup_supp}
\end{figure}

\begin{figure}[t]
    \centering
    \includegraphics[width=.95\columnwidth]{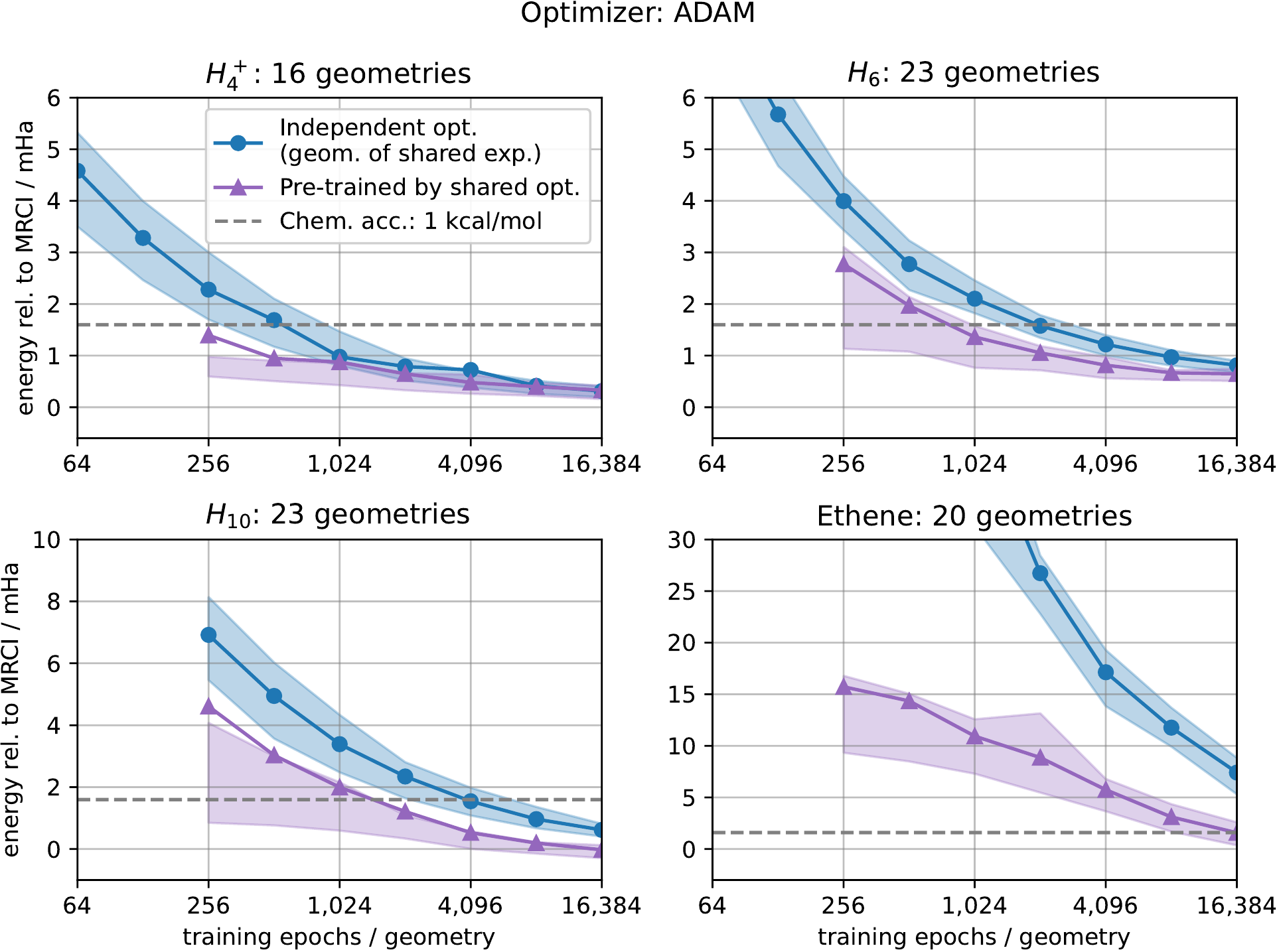}
    \caption{Mean evaluation error relative to a reference calculation (MRCI-F12(Q)/cc-pVQZ-F12 (Molpro)) as a function of training epochs per geometry for four different sets of nuclear geometries. Plots show energy errors for pre-training on different geometries of the same molecules compared to a standard non-pretrained optimization. Shadings indicate the area between the 25\,\% and the 75\,\% percentile of considered geometries. For each method, we plot intermediary and final results for independent optimizations that ran for a total number of \numprint{16384} epochs.}
    \label{fig:reuse_speedup_supp}
\end{figure}

Figure~\ref{fig:parallel_speedup_supp} and Figure~\ref{fig:reuse_speedup_supp} show results for the experiments presented in the main text in Section~2.1 and Section~2.2, respectively, when using the first-order ADAM \cite{Adam} optimizer instead of second-order K-FAC optimization. The changed optimization hyperparameters are compiled in Table~\ref{table:hyperparameters_supp}.

\begin{table}[b]
\caption{Optimization hyperparameters for DeepErwin for the experiments reported in the supplementary materials. Changes to the hyperparameters in the main text are written in boldface.\label{table:hyperparams}}

\begin{tabularx}{\textwidth}{|X|X|c|}
\hline
\multirow{2}{*}{ {\textbf{Optimization general}}}
		 &Optimizer		                        & \textbf{ADAM}\\ \cline{2-3}
		 &Batch size 	                        & $512$	\\ \cline{2-3}
\hline
\hline
\multirow{2}{*}{ {\textbf{Independent optimization}}} 		 &Initial learning rate $\text{lr}_0$	& $\mathbf{1.5 \times 10^{-3}}$	\\ \cline{2-3}
		 &Learning rate decay   	            & $\text{lr}(t) = \frac{\text{lr}_0}{1+t/1000}$ 	\\ \cline{2-3}
\cline{2-3}
\hline
\hline
\multirow{4}{*}{ \shortstack[l]{\textbf{Shared optimization}\\  {\textbf{(Fig.~\ref{fig:parallel_speedup_supp})}}}}
		 &{$\text{lr}_0$} for non-shared weights 	                    & {$\mathbf{1.5 \times 10^{-3}}$} 	\\ \cline{2-3}
		 &{$\text{lr}_0$} for shared weights	                & {$\mathbf{1.5 \times 10^{-3}}$}	\\ \cline{2-3}
		 &{Learning rate decay}	                & { $\text{lr}(t) = \frac{\text{lr}_0}{1+t/10000}$}	\\ \cline{2-3}
		 &Training scheduler	                & Std. deviation	\\ \cline{2-3}
\hline
\hline
\multirow{3}{*}{\shortstack[l]{{\textbf{Optimization after pre-training}}\\ {\textbf{(Fig.~\ref{fig:reuse_speedup_supp})}}}}
		 &{$\text{lr}_0$ for non-shared pre-trained weights} 	                    & $\mathbf{1.5 \times 10^{-3}}$\\ \cline{2-3}
		 &{$\text{lr}_0$ for shared pre-trained weights}	                & $\mathbf{1.5 \times 10^{-3}}$	\\ \cline{2-3}
		 &{Learning rate decay}	                & { $\text{lr}(t) = \frac{\text{lr}_0}{1+t/1000}$}	\\ \cline{2-3}
\hline
\end{tabularx}
\label{table:hyperparameters_supp}
\end{table}
\FloatBarrier
\subsection{Independent optimization as pre-training}
\begin{figure}[h]
    \centering
    \includegraphics[width=.8\columnwidth]{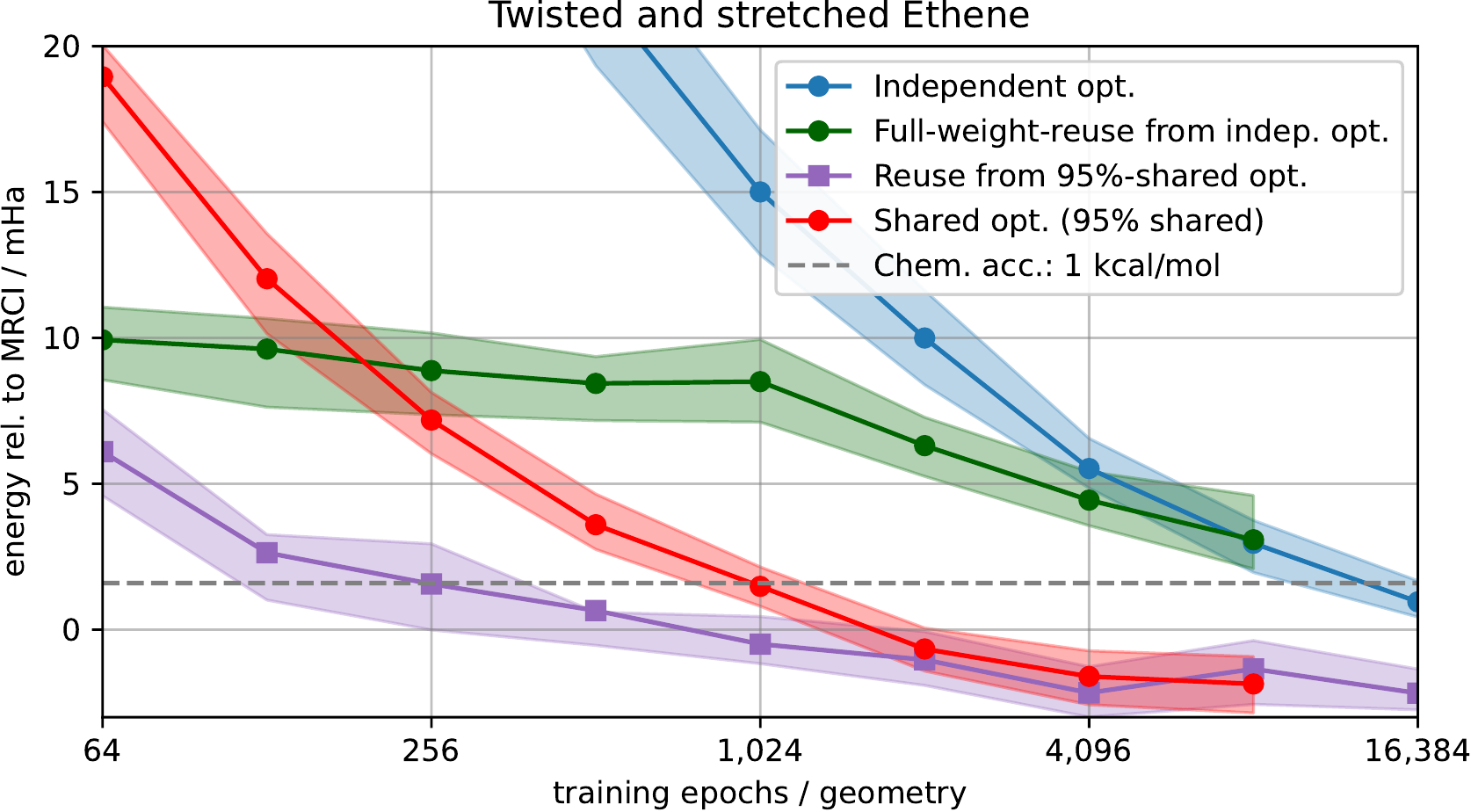}
    \caption{Mean evaluation error relative to the reference calculation (MRCI-F12(Q)/cc-pVQZ-F12) as a function of training epochs per geometry for a set of differently twisted and streteched ethene configurations. Shadings indicate the area between the 25\,\% and the 75\,\% percentile of considered geometries. For each method, we plot intermediary and final results for optimizations that ran for a total number of \numprint{16384}, respectively \numprint{8192}, epochs per geometry.}
    \label{fig:reuse_from_indep_pretraining}
\end{figure}

In the main text, mere independent optimization was used as a benchmark for all numerical experiments. Here, we consider an additional baseline, for which independent optimizations for different ethene configurations were fully initialized with weights from a wavefunction that had already been optimized for a similar but different molecular configuration using an independent optimization scheme. Optimization hyperparameters for the new benchmark are the same as the ones reported in Table~1 in the main text. The respective results, including a comparison with shared optimization and independent optimzation that was pre-trained on a set of different ethene configurations using shared-weight constraints, are shown in Figure~\ref{fig:reuse_from_indep_pretraining}. While the new benchmark seems to be advantageous during early optimization as compared to a scheme that applies a weight-sharing constraint for 95\,\% of the weights in the model, at the time the wavefunctions reach chemical accuracy, shared optimization without pre-training outperforms this new baseline to a degree comparable with the results shown in Figure~2 in the main text.
\printbibliography